\documentclass[conference]{IEEEtran}

\usepackage{amsmath,amsfonts}
\usepackage{algorithmic}
\usepackage{array}
\usepackage[caption=false,font=normalsize,labelfont={sf},textfont={sf}]{subfig}
\usepackage{textcomp}
\usepackage{stfloats}
\usepackage{url}
\usepackage{verbatim}
\usepackage{graphicx}
\def\BibTeX{{\rm B\kern-.05em{\sc i\kern-.025em b}\kern-.08em
    T\kern-.1667em\lower.7ex\hbox{E}\kern-.125emX}}
\usepackage{balance}

\usepackage{
    hyperref,
    color,
    multirow,
    arydshln,
    booktabs,
}

\usepackage[table]{xcolor}

\newcommand\first[1]{\textbf{#1}}  
\newcommand\second[1]{\underline{\smash{#1}}}  
\newcommand\colname[1]{\textbf{#1}}  
\newcommand\example[1]{\textit{``#1''}}  
\newcommand\na{\raisebox{0.02cm}{\scriptsize N/A}}
\newcommand\mytexttt[1]{``{\upshape \texttt{#1}}''}

\newcommand\mytitle{{CoNeTTE: An efficient Audio Captioning system leveraging multiple datasets with Task Embedding}}

\begin{document}

\title{\mytitle}


\author{Étienne Labb\'e, Thomas Pellegrini, Julien Pinquier \\
    IRIT, Université de Toulouse, CNRS, Toulouse INP, UT3, Toulouse, France \\
}

\markboth{IEEE/ACM Transactions on Audio, Speech, and Language Processing}%
{Labb\'e \MakeLowercase{\textit{et al.}}: \mytitle}

\maketitle

\begin{abstract}

Automated Audio Captioning (AAC) involves generating natural language descriptions of audio content, using encoder-decoder architectures. An audio encoder produces audio embeddings fed to a decoder, usually a Transformer decoder, for caption generation. In this work, we describe our model, which novelty, compared to existing models, lies in the use of a ConvNeXt architecture as audio encoder, adapted from the vision domain to audio classification. This model, called CNext-trans, achieved state-of-the-art scores on the AudioCaps (AC) dataset and performed competitively on Clotho (CL), while using four to forty times fewer parameters than existing models. We examine potential biases in the AC dataset due to its origin from AudioSet by investigating unbiased encoder's impact on performance. Using the well-known PANN's CNN14, for instance, as an unbiased encoder, we observed a 1.7\% absolute reduction in SPIDEr score (where higher scores indicate better performance). To improve cross-dataset performance, we conducted experiments by combining multiple AAC datasets (AC, CL, MACS, WavCaps) for training. Although this strategy enhanced overall model performance across datasets, it still fell short compared to models trained specifically on a single target dataset, indicating the absence of a one-size-fits-all model. To mitigate performance gaps between datasets, we introduced a Task Embedding (TE) token, allowing the model to identify the source dataset for each input sample. We provide insights into the impact of these TEs on both the form (words) and content (sound event types) of the generated captions. The resulting model, named CoNeTTE, an unbiased CNext-trans model enriched with dataset-specific Task Embeddings, achieved SPIDEr scores of 44.1\% and 30.5\% on AC and CL, respectively. For the sake of reproducibility, we have made our code publicly available\footnote{\url{https://github.com/Labbeti/conette-audio-captioning}}.


\end{abstract}


\section{Introduction}

\IEEEPARstart{I}{N} recent years, natural language processing has gained significant popularity and emerged as a universal interface facilitating human-machine interactions, owing to remarkable advancements in machine learning systems. Unlike predefined class sets, human-generated free text typically contains more comprehensive information, encompassing relationships between entities, complex scene descriptions, and object attributes. Initially introduced for image content, the image captioning task aims to create models capable of describing visual content using natural language (e.g.,~\cite{7298935}). This concept was extended to audio processing, giving rise to Automated Audio Captioning (AAC) (e.g.,~\cite{drossos_2017_waspaa}), with the goal of generating text-based descriptions for audio content. 


AAC systems use encoder-decoder architectures, where an audio encoder provides a sequence of embeddings to a decoder, either a recurrent neural network or a Transformer decoder, responsible for generating a caption~\cite{Mei2022,Xu2022ACS}. In this work, we describe our model referred to as CNext-trans, in which the audio encoder is a fully convolutional neural network adapted from the vision domain, the well-known ConvNeXt architecture~\cite{liu2022convnet}. The decoder is a vanilla Transformer decoder, trained from scratch on the AAC datasets at hand. As we will report here-after, CNext-trans performed very favorably on AudioCaps (AC)~\cite{kim_etal_2019_audiocaps} and Clotho (CL)~\cite{drossos_clotho_2019}, two datasets widely used in AAC.  

All the existing systems, including ours, use audio encoders pretrained on at least AudioSet (AS)~\cite{gemmeke_audio_2017}. This rises a bias issue when doing experiments on AC, since the AC audio recordings were selected from the training subset of AS. In this work, we compare the use of biased and unbiased audio encoders on AC, and we observe that the difference in AAC performance between the two is relatively small.  

Next, we explore combining four AAC datasets (AC, CL, MACS~\cite{Martin2021b}, WavCaps~\cite{mei2023WavCaps}) for training. This was motivated by the fact that a model trained on AC performs badly on CL, and vice versa. We observe that simply combining all these data enhanced the overall performance across datasets, but it still fell short compared to models trained on a single target dataset, indicating the absence of a one-size-fits-all model. To bridge the performance gap between datasets, we introduce Task Embedding (TE) tokens at the input of the decoder, enabling the model to identify the dataset source for each input sample. We analyze the influence of these TEs on the form (words) and content (sound event types) of generated captions. Our final CoNeTTE model, an unbiased CNext-trans with dataset-specific Task Embeddings, achieved the best cross-dataset performance trade-off, and is a path towards a one-size-fits-all model.



After presenting recent AAC works from the literature, we describe our system in Section~\ref{sec_our_system} and the datasets used to train and evaluate it with the different potential bias problems in Section~\ref{sec_datasets_and_biases}. We detail all the hyperparameters, metrics and first audio tagging results to study data biases in Section~\ref{sec_exptal_setup}. Then, we present our final results on AC and CL, and compare to state-of-the-art systems. Finally, we report dataset combining experiments and the introduction of our task embedding strategy in Section~\ref{ssec_with_external_data}.

\section{Related Work}
\label{sec_related_works}

To address the AAC task, numerous architectures, data augmentation techniques, and training objectives have been proposed in the literature. In this study, we focus on the methods applied to the two main datasets used in our article: AC and CL. We exclude works that use a Reinforcement Learning procedure to artificially boost captioning performance metrics, as they tend to produce low-quality captions with repetitive n-grams~\cite{aac_rl_degenerated}. Additionally, ensemble learning results, which combine several model predictions, have also been excluded, as they are commonly employed in AAC challenges.

All recent systems employ a pretrained audio encoder network, combined with a Transformer-like architecture~\cite{NIPS2017_3f5ee243} as a word decoder. The authors of~\cite{aac_multi_tta} employed a test-time data augmentation technique named MM-TTA~\cite{mm_tta} to average the representations of multiple augmented versions of the same input. Gaussian noise and SpecAugment~\cite{Park_2019} were used to produce variants, and the method can average the representations at different levels in the network. The system also utilizes mixup~\cite{zhang2018mixup} data augmentation during training between audio waveforms and spectrograms and concatenates the corresponding caption labels. They employed a full Transformer architecture with an encoder network named Audio Captioning Transformer~\cite{mei2021audio}, pretrained on AS for Audio Tagging like most models. Their system currently holds the state-of-the-art position on AC without external training data.

Some studies have proposed using a pretrained decoder to improve the language generation part, as seen in~\cite{Gontier2021}, where the authors employed a Transformer decoder named BART~\cite{lewis_etal_2020_bart} to enhance caption quality. However, the audio embedding space is usually too different from the sentence embedding space, leading to the pretrained decoder forgetting most of its previous knowledge. To overcome this, they combined two audio encoders. The first one produces sound event tags detected in the audio file, which are given to the BART input embedding layer and added to the audio embeddings extracted from another audio encoder. This approach is expected to make the BART inputs closer to the ones expected by the pretrained weights. The first encoder is a YAMNet~\cite{yamnet} architecture, and the second encoder is the \texttt{Wavegram-Logmel-CNN} from the Pretrained Audio Neural Networks study (PANN)~\cite{kong_panns_2020}.

Recently, the authors of~\cite{mei2023WavCaps} introduced a new captioning dataset named WavCaps. Each caption in this dataset is a raw description crawled from various websites and processed with ChatGPT\footnote{\url{https://openai.com/blog/chatgpt/}}. Their dataset is an order of magnitude larger than other AAC datasets. They trained several models for different audio-language tasks, comparing CNN14 and HTS-AT~\cite{htsat} as pretrained audio encoders, and BART as a pretrained decoder. HTS-AT-BART achieved the current state-of-the-art score on AC with the use of external data.

On CL, the DCASE challenges compile most of the results obtained over the last years\footnote{\url{https://dcase.community/}}. The authors of~\cite{Han2021} proposed using a ResNet38 model from PANN as the encoder, with a standard Transformer decoder model. They also crawled audio files from the Web and processed uploaders' raw descriptions to create a larger training corpus for their model, obtaining data from several websites. Additionally, some data augmentation techniques, such as noise and reverberation, were used to improve generalization.

The PaSST-trans~\cite{kouzelis2022_t6a} system proposed using another pretrained encoder named PaSST~\cite{koutini22_interspeech}. They used a Transformer model with the Patchout method, randomly dropping patches of the spectrograms and flattening the results to obtain shorter sequences as input. Their models were trained on AC, CL, and MACS, with the use of mixup and label smoothing.

The authors of~\cite{won2021_t6} presented a simple model based on CNN14 and a Transformer decoder, trained exclusively on CL, which makes this system similar to our CNN14-trans (see below). Three differences compared to our variant may explain their better results: they employed Stochastic Weight Averaging~\cite{izmailov2018averaging} to average models over several epochs for testing (similar to ensembling), they pretrained the word embeddings, and their decoder has a lower number of trainable parameters.

Finally, the system that achieved first place in the most recent DCASE 2023 Challenge Task 6a\footnote{\url{https://dcase.community/challenge2023/}} is the BEATs+Conformer~\cite{wu2023_t6a} system. It utilizes BEATs~\cite{chen2022beats} as a pretrained encoder to produce audio features, which ranks among the current state-of-the-art audio tagging models on AS. The audio features are then downsampled by a trainable Conformer network~\cite{gulati2020conformer} and given to a BART pretrained decoder to generate the captions. Moreover, they used an Instructor-XL model~\cite{su2023embedder}, based on a T5 Transformer~\cite{transformer_t5}, to generate embeddings from captions and used them as targets with Conformer outputs and InfoNCE~\cite{Oord2018Representation} loss. Finally, these outputs are used to feed a pretrained BART decoder to generate the captions.

Our AAC system is similar to the ones described above, with two main differences: i) our encoder is a ConvNeXt model, ii) our decoder is a vanilla Transformer decoder randomly initialized. As described in details here-after, this combination allowed us to achieve high performance, while significantly reducing model size compared to many existing AAC systems.

\section{Our system: CNext-trans}
\label{sec_our_system}

\subsection{Architecture}
\begin{figure*}[ht!]
    \centering
    \includegraphics[width=1.0\linewidth]{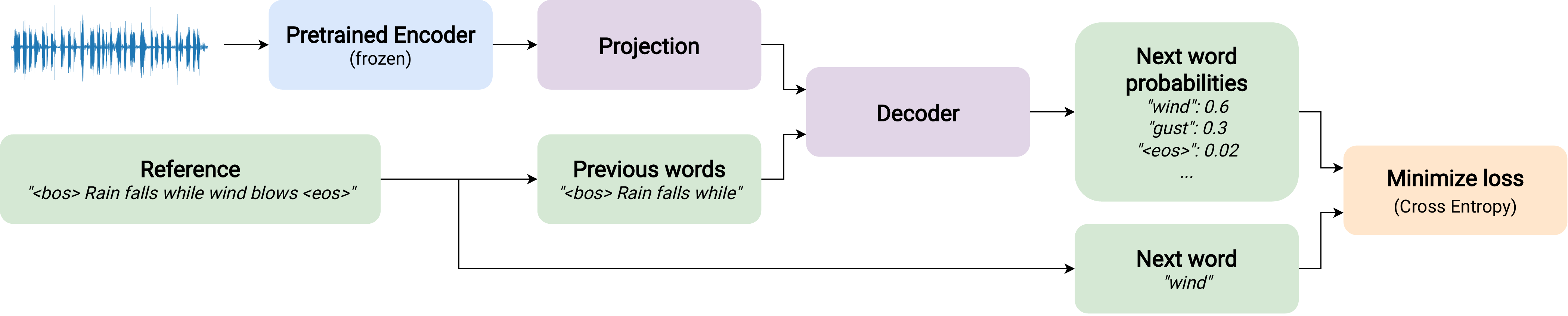}
    \caption{Overview of our AAC training process. The model is composed by an audio encoder which produces a frame-level audio representation of the audio and a captioning decoder which produces the next word distribution according to the previous words and to the audio. This process is repeated for each word in the reference, except for the Begin-Of-Sentence (\texttt{<bos>}) token.}
    \label{fig_aac_system}
\end{figure*}

Our system is a deep neural network that employs a standard encoder-decoder architecture. The encoder part produces frame-level audio embeddings, and the decoder predicts the next word according to the previous ones and to the audio representation. Each caption is preceded by a Begin-Of-Sentence token (\texttt{<bos>}) and followed by an End-Of-Sentence token (\texttt{<eos>}). We train our model using the teacher forcing method, which always gives the ground truth previous tokens to the model, in contrast to scheduled sampling methods which use the previous predicted tokens as inputs to the decoder. The model outputs give the probabilities of the next word that are compared with the ground truth next word using a standard Cross-Entropy (CE) loss. During inference, we loop on the decoder forward method to produce each word of the sentence sequentially by adding the most probable next word to the previous words until an EOS token or the maximal number of words is reached. An illustration of the training procedure is giving in Fig.~\ref{fig_aac_system}.

\subsection{Audio encoder: CNN14 and ConvNeXt-Tiny}

In our experiments, we considered two types of pretrained encoders for AAC: the CNN14 model from PANN and ConvNeXt~\cite{liu2022convnet} (CNext), a computer vision convolutional neural network that we previously adapted for audio tagging in our work~\cite{pellegrini2023adapting}. We observed that a superior encoder generally results in better audio representations, which are crucial for the decoder to produce accurate captions. Both models were pretrained on AS, the largest audio tagging corpus with approximately 2 million labeled audio files. The code and weights of our CNext audio encoder are available\footnote{\url{https://github.com/topel/audioset-convnext-inf}}.

CNN14 is a standard vanilla convolutional-based network consisting of six ConvBlock layers. Each ConvBlock comprises two sequences of convolutions, batch normalization, and ReLU activation, followed by 2-dimensional average pooling. These layers generate a frame-level sequence of embeddings with a size of 2048. We removed the pooling, projection, and classification layers used to predict the AS classes.

The CNext models are based on depthwise separable convolutions~\cite{Chollet_2017_CVPR} (DSC) and inverted bottleneck~\cite{sandler2018mobilenetv2} (IB) layers. DSC involves a sequence of depthwise convolutions that process feature channels separately, followed by a pointwise convolution to mix them. This technique aims to produce results similar to standard convolutional layers while reducing the number of operations to speed up training and mitigate overfitting. The IB layer is a sequence of a pointwise convolution layer that increases the number of channels, followed by a GELU~\cite{gelu} activation, and by another pointwise convolution layer that reverts the number of channels back to its value at the bottleneck input. The output is then added to the original input to create a residual connection, preventing vanishing gradient issues and reducing the number of parameters compared to a standard residual block. All the hyperparameters and training details of the ConvNeXt-Tiny model that we used are described in~\cite{pellegrini2023adapting}.

We incorporated a small network on top of these encoders to project the audio frame embeddings onto a subspace with the same dimension $d_{model}$ as that of the decoder network. This additional network consists of a 0.5 dropout, a linear layer, a ReLU activation function, and another 0.5 dropout. The linear layer takes embeddings of size 2048 for CNN14 and 768 for CNext.

\subsection{Word decoder}
The decoder network is a standard Transformer decoder~\cite{NIPS2017_3f5ee243} with 6 layers, 8 attention heads per layer, a global embedding size $d_{model}$ set to 256, a layer norm epsilon set to $10^{-5}$, a global dropout probability of 0.2 and a feedforward dimension size set to 2048. We used the GELU activation layer in the decoder. These hyperparameters were obtained after a few tests on both data sets.

Like almost all AAC systems, we employed the well-known beam-search algorithm widely used in speech processing~\cite{beam_search_1976} during inference to improve subtitle quality and accuracy. We implemented a per-batch version of the beam search algorithm which speeds up the sentence generation by a factor of 10, but requires more memory to run. This algorithm usually leads to generic and repetitive content, which can be penalized by certain metrics. We introduced a constraint during inference to avoid repeats to words that are in previous tokens. We decided to allow repeating a pre-defined list of stop-words from the Natural Language ToolKit (NLTK)~\cite{bird2009natural} package. We found that it can improve diversity and limit repetition while slightly decrease n-gram based metrics. This constraint can forbid a lot of repetitions in the predicted sentences like \example{a man speaks while a man speaks}, but we found that the model can still produce repetitive content with closely related words like in \example{children speak and child speaks}. In addition, we limit the minimal number of tokens predicted to 3 and the maximal number to 20 or 30 during inference to avoid few cases of degenerated sentences.

\subsection{Tackling overfitting}

We noticed in our first experiments that our model can easily overfit the training data, even with small networks with less than 10M parameters. We found that using a large weight decay value with the AdamW optimizer~\cite{adamw} drastically overcame this issue~\cite{labbé2023multitask}. In addition, we searched to apply data augmentation during training. This case is particular for AAC task since a lot of audio transformations (reverb, background noise, pitch shifting...) of the audio can be described in the target captions. We decided to focus on the following three data-agnostic augmentations.
%
%

Mixup~\cite{zhang2018mixup} is used on the decoder inputs (audio embeddings and previous token embeddings) as in~\cite{kouzelis2022_t6a,takeuchi2020effects} to improve the robustness of our model. We also tried to mix target labels, but it did not bring any improvements. The algorithm is shown in Eq.~\ref{eq_mixup_aac} and resumes how mixup is used in our system. $x_1$ corresponds to an audio embedding with its label $y_1$  and $x_2$ is another audio embedding from the current batch, with its label $y_2$. $\alpha$ is a fixed hyperparameter, $W$ denotes the input word embedding layer and $f$ is the rest of the AAC decoder network.


\begin{equation}
\begin{split}
    \lambda & \sim \text{Beta}(\alpha, \alpha) \\
    \lambda & = \max(\lambda, 1 - \lambda) \\
    x_{mix} & = \lambda x_1 + (1 - \lambda) x_2 \\
    w_1 & = W(y_{1,prev}) \\
    w_2 & = W(y_{2,prev}) \\
    w_{mix} & = \lambda w_1 + (1 - \lambda) w_2 \\
    z_{mix} & = f(x_{mix}, w_{mix}) \\
    \mathcal{L} & = \text{CE}(z_{mix}, y_{1,next})
    \label{eq_mixup_aac}
\end{split}
\end{equation}

SpecAugment~\cite{Park_2019} is applied on the audio frame embeddings, outputted by the audio encoder. We found that using this augmentation on spectrograms or audio embeddings provides similar improvements, but applying it on embeddings allowed us to pre-compute the encoder outputs and drastically accelerate the experiments. We modified the behavior of this augmentation to mask a proportion of the time and feature axes instead of using an absolute mask size. Each axis is masked twice, with a mask size sampled between 0 and 10\% of the total axis size (number of time steps or embedding size).

Label smoothing~\cite{szegedy2015rethinking} is employed on target captions to limit the maximum probability of the model for each token and reduce overfitting.

\section{Datasets and potential biases}
\label{sec_datasets_and_biases}

\subsection{Descriptions}
In a first series of experiments, we used AC and CL separately. In a second setting, we added training data from Multi-Annotator Captioned Soundscapes (MACS) and WavCaps. We report statistics about the number of word types, tokens, etc. about the training subsets in Table~\ref{tab_datasets}, to show how these datasets differ in content.

AudioCaps (AC) is the largest human-labeled AAC dataset of originally 51,308 audio files, taken from a subset of the (unbalanced) training subset of AudioSet (AS). AC contains three splits, and because the original YouTube videos are removed for various reasons, our version contains only 46,213 out of 49,838 files in the training subset, 464 out of 495 in the validation subset and 912 out of 975 files in the test subset. The training subset files are described by only one caption per audio, while the validation and test files are described by five captions each.

Clotho (CL) is a smaller dataset containing files extracted from the FreeSound website. The training subset contains 3840 files and the validation and test subsets contain 1045 files each. Unlike some papers in the literature~\cite{mei2022diverse}, we did not use the validation subset to train our model. All subsets contain five captions per audio and each caption was corrected by a second set of annotators to remove grammatical, subjectivity, fluency, or repetition errors.

Multi-Annotator Captioned Soundscapes (MA) is another AAC dataset containing audio files from the development subset of the TAU Urban Acoustic Scenes 2019~\cite{Mesaros2018_DCASE} dataset. The sound events have been recorded in different acoustic scenes like airports, public squares, and parks. Each annotator labeled audio files with a predefined set of sound event classes, then with a free-text caption.

WavCaps (WC) is the largest audio captioning dataset, containing 403,050 audio-caption pairs. It is a collection of audio recordings from the AudioSet Strongly Labeled subset, FreeSound, SoundBible and BBC Sound Effects. Captions have been generated by a ChatGPT model, using the audio event classes for the AudioSet subset or the original human descriptions for the other subsets. WC features a large diversity in audio length: from 1 second to 18 hours long.

We used our own source code to download and load the dataset files, available as a Python package named aac-datasets\footnote{\url{https://github.com/Labbeti/aac-datasets}}. 

\begin{table}[ht]
    \centering
    \caption{Statistics for AC, CL, MA and WC subsets used for training.}
    \label{tab_datasets}

    \begin{tabular}{|l|c|c|c|c|}
        \hline
        & \colname{AC} & \colname{CL} & \colname{MA} & \colname{WC} \\
        \hline
        Sample rate (Hz) & 32000 & 44100 & 48000 & 32000 \\
        Audio duration range (s) & 0.5-10 & 15-30 & 10 & 1-67109 \\
        Nb audio & 46,213 & 3839 & 3930 & 403,050 \\  
        Audio size (h) & 126.6 & 24.0 & 10.9 & 7563.3 \\  
        Vocabulary size & 4585 & 4369 & 2721 & 24600 \\  
        Nb words & 401,650 & 217,360 & 159,879 & 3,161,823 \\  
        Caption length range & 2-40 & 8-20 & 2-40 & 2-38 \\  
        Caption length mean & 8.7 & 11.3 & 9.3 & 7.8 \\
        Nb captions per audio & 1 & 5 & 2-5 & 1 \\
        \hline
    \end{tabular}
\end{table}

\subsection{Potential training data biases in AAC}

We identified two types of potential biases when carrying out AAC experiments on AC and CL, which concern either the pretrained audio encoders or the whole AAC systems. These biases may arise because of overlaps within the data sources, either between AAC and Audio Tagging (AT) datasets, or between AAC datasets. We report the overlap proportions between datasets CL, AC, WC, AS and FSD50K in Table~\ref{tab_overlap_data}. 

\begin{figure}[!htb]
    \centering
    \includegraphics[width=1.0\linewidth]{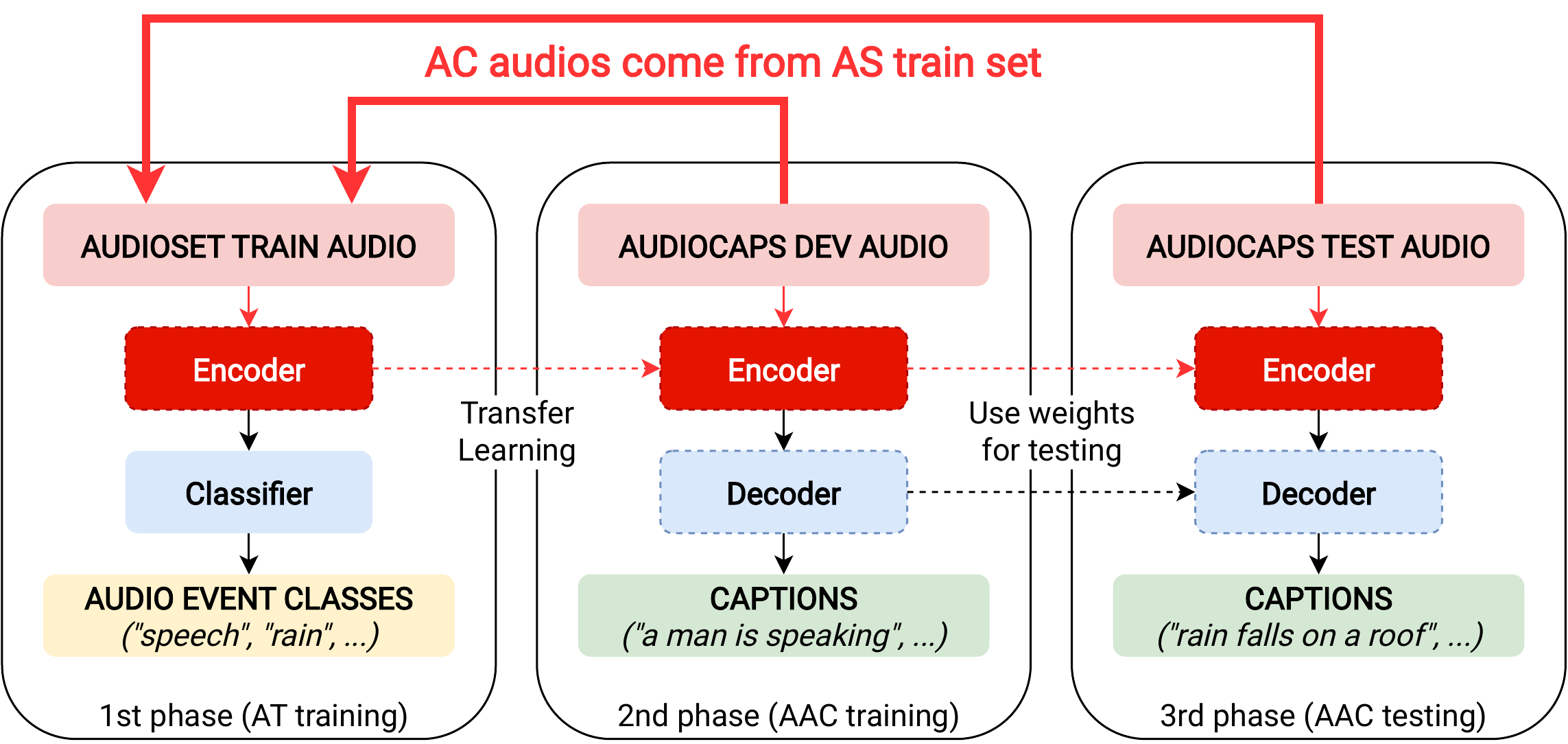}
    \caption{Illustration of the bias that may occur when using an audio encoder pretrained on AS for experiments on AC. All the subsets of AC come from the AS training subset, which means that the encoder part of the AAC systems were trained on the validation and testing data for audio tagging, a task closely related to AAC.}
    \label{fig_audiocaps_bias}
\end{figure}

Most AAC systems use audio encoders pretrained to perform AT, either on AS or on the audio tagging dataset FSD50K~\cite{fonseca2022fsd50k}. However, the whole AC dataset is actually a subset of the AS training subset, which means that an encoder pretrained on AS has already seen 100\% of the audio files of AC, even the validation and testing ones. This implies that the encoder already ``knows'' the sound events of AC. This bias concerns all audio-language tasks (audio captioning and audio retrieval) involving AS and AC in their procedure. Figure~\ref{fig_audiocaps_bias} summarizes this data bias (or data leak) that we want to highlight.

To a much smaller extent, the same problem occurs when pretraining on FSD50K and running AAC experiments on CL, since about 5\% of the CL files are shared with FSD50K's training subset (extracted from the FreeSound Website).

The second type of bias, which may impact whole AAC systems, involves the WC dataset. WC comprises files from both FreeSound and AS, which are the data sources of CL and AC, respectively. One has to take care of removing these overlaps (18\% in common with AC, and 89\% in common with CL), when pretraining an audio encoder (first bias mentioned above), and when training an AAC system.

There could also be overlaps between FSD50K or WC with the CL private subsets used in the DCASE challenge task 6a, which means that the results of the models trained on these datasets could be overestimated. These potential overlaps remain unknown to us, since the audio file IDs are not available.

\begin{table}[hb]
    \centering
    \caption{Overlap proportions between AAC/AT datasets.}
    \label{tab_overlap_data}

    \begin{tabular}{|c|c|c|c|}
        \hline
       \colname{Dataset A} & \colname{Dataset B} & \colname{Overlap (\%)} & \colname{B labels type} \\
        \hline
        AC & AS-train & 100.0 & Tags \\
        CL & FSD50K-train & \hphantom{00}5.4 & Tags \\
        AC & WC & \hphantom{0}17.6 & Captions \\
        CL & WC & \hphantom{0}89.0 & Captions \\
        \hline
    \end{tabular}
\end{table}


\subsection{Task Embedding}
\label{ssec_task_emb}

After obtaining our results on AC and CL without external data, we attempted to improve performance by adding more training data. Surprisingly, we observed a decrease in results. Previous work~\cite{Berg2021} noted that the sound events and writing styles of AC and CL can form distinct domains, which can perturb models trained on both datasets compared to those trained separately. Many AAC studies~\cite{kouzelis2022_t6a,mei2023WavCaps,yuan2021_t6} handle external data by splitting the training into two phases: the first one involves training on all available captioning datasets, while the second one entails fine-tuning on the target dataset only (AC or CL).


However, we believe that a single AAC system should be capable of leveraging multiple captioning datasets, given a hint to the model. To achieve this, we introduced a Task Embedding (TE) tag as input to the decoder, dependent on the audio dataset source, such as \texttt{<bos\_ac>}, \texttt{<bos\_cl>}, etc. This tag replaces the \texttt{<bos>} token used at the sentence's beginning and enables the model to generate an output closer to the desired writing style. We initially employed this method in our participation in the DCASE Challenge 2023~\cite{labbe2023_t6a}, where our best system achieved third place. In this study, we expanded on this approach by using more datasets and TE tokens. Interestingly, another DCASE participant~\cite{kadlcik_t6a} also utilized a TE approach to leverage different datasets (CL, AC, and AS), employing a system based on the speech automatic recognition system Whisper~\cite{radford2023robust}. However, their system performed worse and was ranked seventh in the challenge.

To test this method, we attempted to add all publicly available AAC datasets: AC, CL, MA, and WC. We excluded overlapping data between AC and WC and between CL and WC to avoid biases. Additionally, we removed audio files lasting for more than 30 seconds in WC, as this is the maximum length of the test audio files of CL. As WC contains four different sources, we added a task embedding tag for each of them, resulting in seven tasks for all datasets. The concatenation of all the audio captioning datasets (CL+AC+MA+WC) yielded 316,122 training audio files. Moreover, we decided to balance data with the target dataset (AC or CL). For example, on CL, we took the 3,840 files of CL and randomly selected another 3,840 files from the other datasets, resulting in 7,680 audio training seen files per epoch. For AC, we used their 46,213 files and an equal number from other datasets to train with 92,426 files per epoch. We tested other balancing strategies, but this approach yielded the best results.


\section{Experimental setup}
\label{sec_exptal_setup}

\subsection{Data pre-processing}

All the captions are written in lowercase, and all punctuation characters are removed. Sentences are tokenized using spaCy~\cite{spacy}. For CL and MA, we resample all audio files from 44.1 kHz and 48 kHz to 32 kHz to match the input sampling rates of the encoders. During training, for CL, we randomly select one of the five captions for each audio file at each epoch in the training subset.

As for AC, we manually corrected a portion of the training captions. This subset contains various mistakes, such as typographic errors (e.g., \example{Continous} corrected to \example{Continuous}), named entities (e.g., \example{Michael Jackson}), invalid descriptions (e.g., \example{Video is unplayable}), speech content (e.g., \example{A large crowd of people chanting “USA” [...]}), or grammar errors (e.g., \example{A engine} corrected to \example{An engine}). Additionally, we excluded captions containing more than 40 words. To create a second version of the AC training subset, we manually fixed 968 captions and deleted 28 files.

\subsection{Metrics}
Evaluating AAC system outputs is a challenging task, as the predicted candidates should contain the same audio events as the references (ground truth), but not necessarily described in the exact same way. Historically, machine translation metrics were used for system comparison, but they primarily rely on n-gram overlapping, which diminishes their correlation with human judgments in captioning evaluation.

The CIDEr-D~\cite{vedantam_cider_2015} metric was developed for image captioning and takes word frequency into account. It calculates the cosine similarity of the TF-IDF scores for common n-grams in both candidates and references. SPICE~\cite{anderson_spice_2016} proposes a comparison of scene graphs containing semantic propositions extracted from sentences using a parser, handcrafted grammar, and custom rules. However, CIDEr-D tends to overvalue candidates containing infrequent n-grams, even if they lack syntactic correctness~\cite{aac_rl_degenerated}, while SPICE tends to give zero scores when the extracted propositions do not match~\cite{zhou_can_2022}. SPIDEr~\cite{liu_improved_2017} was introduced to consider both metrics by averaging their scores. Nevertheless, as CIDEr-D values range from 0 to 10 and SPICE values range from 0 to 1, CIDEr-D has a more significant influence on the SPIDEr score.

In several recent studies, SPIDEr  has faced criticism, for example for its lack of sensitivity to repetitions~\cite{Martin_Morato2022_2,spice_plus,labbé2022automatic}. To establish a more robust evaluation of AAC systems, the authors of~\cite{zhou_can_2022} created FENSE, a metric based on two pretrained models for comparing sentence embeddings rather than n-grams. FENSE employs two models: a Sentence-BERT (SBERT) model and a Fluency Error Detector model. The SBERT model is trained to generate fixed-size embeddings that can be compared using cosine similarity (referred to as SBERT-sim). The Fluency Error Detector model is trained to detect common mistakes made by AAC systems, such as repetitive n-grams, incomplete sentences, repetitive adverbs, missing conjunctions, and missing verbs. The FENSE score is equal to SBERT-sim when no error is detected by the Fluency Error Detector model; otherwise, the score is divided by 10.

In this work, we report SPIDEr, FENSE and the number of unique words used (named {\#Words} or sometimes {Uniq.} in the literature). We have regrouped all the code needed to compute these metrics into our package aac-metrics\footnote{\url{https://github.com/Labbeti/aac-metrics}}, which is publicly available online.

\begin{table}[htbp]
    \centering
    \caption{Training and decoding hyperparameters per dataset.}
    \label{tab_hparams}

    \begin{tabular}{|l|cc|}
        \hline
        \multirow{2}{*}{\colname{Name}} & \multicolumn{2}{c|}{\colname{Value}} \\
        & \colname{AC} & \colname{CL} \\
        \hline
        Batch size & \multicolumn{2}{c|}{512} \\
        Optimizer & \multicolumn{2}{c|}{AdamW} \\
        Initial learning rate ($\text{lr}_0$) & \multicolumn{2}{c|}{$5 \cdot 10^{-4}$} \\
        $\beta_1$ & \multicolumn{2}{c|}{$0.9$} \\
        $\beta_2$ & \multicolumn{2}{c|}{$0.999$} \\
        $\epsilon$ & \multicolumn{2}{c|}{$10^{-8}$} \\
        Weight decay & \multicolumn{2}{c|}{$2$} \\
        Gradient clip norm type & \multicolumn{2}{c|}{$\ell^2$} \\
        mixup param. ($\alpha$) & \multicolumn{2}{c|}{$0.4$} \\
        Min prediction size & \multicolumn{2}{c|}{$3$} \\
        \hdashline
        Nb. Epochs ($K$) & 100 & 400 \\
        Gradient clip norm value & 10 & 1 \\
        Label smoothing & 0.1 & 0.2 \\
        Max prediction size & {$30$} & {$20$} \\
        Beam size & $2$ & $3$ \\
        \hline
    \end{tabular}
\end{table}

\subsection{Hyperparameters}
We provide a detailed account of all the training hyperparameter values in Table~\ref{tab_hparams}. Throughout our experiments, we validate and select our models using the FENSE score on the validation subset. Unlike the validation likelihood loss, this metric evaluates the sentence generation of our models, and we have found it to be more stable than the loss or SPIDEr values. The latter tends to vary drastically among different seeds due to its sensitivity to predicted n-grams~\cite{labbé2022automatic}.

The weight decay is not applied to the bias weights of the networks. To prevent models from collapsing, we used gradient clipping~\cite{pascanu2013difficulty}. During the AAC training phase, we freeze the encoder, enabling us to pre-compute audio embeddings and significantly reducing the training time. On a single GPU V100 with 32 GB of memory, AAC experiments take only one hour on AC and three hours on CL. The learning rate is decreased during training at the end of each epoch $k$ using a cosine scheduler rule~\ref{eq_cos_scheduler}:

\begin{equation}
    \label{eq_cos_scheduler}
    \text{lr}_k = \frac{1}{2} \left(1 + \cos \left( \frac{k \pi}{K} \right) \right) \text{lr}_0
\end{equation}


\subsection{Audio tagging results on AudioCaps with and w/o bias}

In this subsection, we evaluate the impact of the training data bias on AC first in terms of audio tagging. We pretrained CNN14 and CNext encoders on AS with and without the AC training, validation, and testing files. Table~\ref{tab_at_results} reports the mean Average Precision (mAP) scores of these classifiers on the AS evaluation subset, as well as on the validation and testing subset files of AC. We compare the CNN14 original weights from PANN with our own trained weights (denoted as {CNN14*}), as well as with our own encoder CNext.

\begin{table}[htbp]
    \centering
    \caption{Tagging mAP scores on AS eval, AC val and AC test.}
    \label{tab_at_results}

    \begin{tabular}{|l|c|c|c|c|}
        \hline
        \colname{Encoder} & \colname{Train data} & \colname{AS-eval} & \colname{AC-val} & \colname{AC-test} \\
        \hline
        CNN14~\cite{kong_panns_2020} & AS & {.431} & {.717} & {.647} \\
        {CNN14*} & AS & {.441} & {.755} & {.727} \\
        CNext & AS & \first{.471} & \first{.774} & \first{.749} \\
        \hdashline
        {CNN14*} & AS-AC & {.434} & {.642} & {.537} \\
        CNext & AS-AC & {.465} & {.669} & {.585} \\
        \hline
    \end{tabular}
\end{table}

The audio tagging results show that, as expected, models trained on the full AS (with bias) achieved significantly higher mAP scores than without bias (lines denoted AS-AC in the table). CNext's mAP on AC-test decreased from 0.749 to 0.585, representing a relative decrease of -21.9\%. This suggests that the model already possesses good knowledge of the sound events in AC-val and AC-test subsets. However, the score on AS-eval remains stable, with only a slight decrease from 0.471 to 0.465 (-1.3\% relative drop). Similar trends are observed with CNN14*, indicating the potential importance of considering this bias when evaluating AAC systems. In the next section, we report AAC results, and we will see that the impact of this specific bias on AC is somehow smaller than in audio tagging, with SPIDEr score relative reductions of about -5\%.

\begin{table*}[ht]
    \centering
    \caption{AAC results on AC and CL testing subsets. Enc$^*$ means that there is a potential bias in the encoder. CNN14 indicates that we used the original weights from PANN, while CNN14* denotes our own trained weights. WC$^{\text{-AC}}$ refers to the WavCaps dataset without sources overlapping with either the AC validation or testing subsets. WC$^{\text{-CL}}$ denotes the WavCaps dataset without sources overlapping with CL validation and testing subsets.}
    \label{tab_aac_results}


    \begin{tabular}{|c|l|c|c|c|c|c|c|c|c|}
        \hline
        \colname{Test} & \multirow{2}{*}{\colname{System}} & \colname{Pretraining} & \colname{Training} & \multirow{2}{*}{\colname{Biased}} & \multirow{2}{*}{\colname{SPIDEr}} & \multirow{2}{*}{\colname{FENSE}} & \multirow{2}{*}{\colname{\#Words}} & \colname{Trainable} & \colname{Frozen} \\
        \colname{data} && \colname{data} &\colname{data} & &&&& \colname{params} & \colname{params} \\
        \hline
        \multirow{9}{*}{AC} & Cross-referencing & $\emptyset$ & $\emptyset$ & {\na} & {.559} & {.680} & {944.0} & {0} & {0} \\
        \cdashline{2-10}
        & HTSAT-BART~\cite{mei2023WavCaps} & AS & AC+WC$^{\text{-AC}}$ & Enc$^*$ & {.485} & \na & \na & {171M} & {0} \\
        & Multi-TTA~\cite{aac_multi_tta} & AS & AC & Enc$^*$ & {.475} & \na & \na & 108M & {0} \\
        & PYB~\cite{Gontier2021} & AS & AC & Enc$^*$ & {.465} & \na & \na & {400M} & {0} \\
        & CNN14-trans (ours) & AS & AC & Enc & .443 & .615 & 383.2 & 12.3M & 75.5M \\
        & CNN14*-trans (ours) & AS & AC & Enc & .456 & .625 & 390.4 & 12.3M & 75.5M \\
        & CNext-trans (ours) & AS & AC & Enc & \first{.495} & {.643} & {393.0} & 12.0M & 28.2M \\
        \cdashline{2-10}
        & CNN14*-trans (ours) & AS-AC & AC & No & .439 & .607 & 354.0 & 12.3M & 75.5M \\
        & CNext-trans (ours) & AS-AC & AC & No & \second{.466} & {.633} & {396.4} & 12.0M & 28.2M \\
        \hline
        \multirow{11}{*}{CL} & Cross-referencing & $\emptyset$ & $\emptyset$ & {\na} & {.567} & {.574} & {1818.2} & {0} & {0} \\
        \cdashline{2-10}
        & BEATs+Conformer~\cite{wu2023_t6a} & AS & CL+AC & No & \first{.326} & \na & \na & 127M & 1.5B \\
        & CNN14-BART~\cite{mei2023WavCaps} & AS & CL+WC$^{\text{-CL}}$ & No & {.310} & \na & \na & {219M} & {0} \\
        & ResNet38-trans~\cite{Han2021} & AS & CL+AC+4 others & \na & {.308} & \na & \na & {81M} & {0} \\
        & PaSST-trans~\cite{kouzelis2022_t6a} & AS & CL+AC+MA & No & {.296} & {.511} & \na & {119M} & {441M} \\
        & CNN14-trans~\cite{won2021_t6} & AS & CL & No & {.285} & {\na} & \na & {8M} & {75.5M} \\
        & CNN14-trans (ours) & AS & CL & No & .265 & .482 & 500.2 & 12.2M & 75.5M \\
        & CNN14*-trans (ours) & AS & CL & No & .274 & .491 & 545.2 & 12.2M & 75.5M \\
        & CNext-trans (ours) & AS & CL & No & .299 & {.512} & {636.8} & 11.9M & 28.2M \\
        & CNN14*-trans (ours) & AS-AC & CL & No & .259 & .477 & 522.6 & 12.2M & 75.5M \\
        & CNext-trans (ours) & AS-AC & CL & No & {.301} & {.516} & {628.2} & 11.9M & 28.2M \\
        \hline
    \end{tabular}
\end{table*}

\section{Audio captioning results on AC and CL}
\label{sec_results}

In this section, we discuss the results of our systems with and without bias, and the impact of using external data in the first subsection. We then explore the effect of adding new training data using TEs or not in the second subsection. Finally, we study the impact of TEs on the generated captions.

\subsection{Results without external data}
\label{ssec_without_external_data}

Table~\ref{tab_aac_results} presents results on the AC and CL datasets. ``Human'' top-line scores are computed by randomly excluding one of the five references for each audio file and using it as a candidate. This provides an inter-agreement score (also called cross-referencing score) between the ground truth references, which is repeated five times and averaged for each subset. These scores offer an idea of the upper bound for each metric. We also report results obtained by state-of-the-art methods presented in section~\ref{sec_related_works}, and our own results. All of our results are averaged over the same five initialization seeds.

On AC, we report results from recent systems from the literature, in which we identified a presumable bias due to their pretrained encoders, since they did not mention that the AC files were removed from AS during the AT pretraining of their encoders. Regarding our systems, the CNext-trans model outperformed CNN14-trans in all contexts, as we expected given its higher performance in audio tagging. Our biased CNext-trans model outperformed the previous state-of-the-art method by 1.0\% absolute, without using any external data and with an order of magnitude fewer trainable parameters. However, the unbiased CNext-trans model, which did not use the AC audio files in pretraining (AS-AC), experienced a 2.9\% absolute points drop, representing only a -5.9\% relative decrease. This indicates that the data bias in this pretrained model still had an impact on performance for the AAC task, related to the AT performance reduction of the model (which was a -21.9\% relative drop).  The FENSE scores seem to be correlated with the SPIDEr values and demonstrate that the CNext encoder provided better representations to the decoder part, both with and without data bias. The number of words used varies from 353 to 396 words, but this is not always correlated with SPIDEr.

On CL, the CNext-trans model continues to outperform other models and achieves a performance close to CNN14-BART, despite having fewer parameters and training data. We have attained the state-of-the-art score for models that do not use any external data. We also observe that the richer representation provided by CNext allows the decoder to produce captions with higher word diversity (545 word types with CNN14-trans compared to 636 with CNext-trans).


Furthermore, we note that our SPIDEr score on AC is close to the cross-reference one with only 6.4\% absolute difference, but the SPIDEr scores on CL remains much lower than the cross-reference score, with a 26.2\% absolute difference. However, the FENSE scores are both close (3.7\% and 5.7\%), indicating that the model produces captions that are semantically close to the references but struggle to find the correct n-grams or propositions. AC is much less diverse than CL in terms of vocabulary (944 and 1818 word types, respectively), and the SPIDEr score is mostly influenced by several predictions that almost fully match the references, resulting in a higher SPIDEr score.

\subsection{Improved results with Task Embedding}
\label{ssec_with_external_data}

To create a general-purpose AAC system, we evaluated different training methods for our CNext-trans model on the two evaluation sets, as shown in Table~\ref{tab_both_results}. The first approach involved training our model separately on each dataset. Then, we attempted to add other datasets and balance the target dataset. Finally, we introduced the TE method to improve overall performance.

When we added all the external data described in~\ref{ssec_task_emb} without TE, the results of our system decreased on both datasets, primarily due to the different writing styles present in the combined data. It became evident that a model trained on one specific dataset performed poorly on the other. For example, a model trained on AC achieved only a 14.6\% SPIDEr score on CL. Similarly, a model trained on all the datasets without TE also performed inadequately, with a 19.1\% SPIDEr score.

However, when we introduced TE, the results of CoNeTTE improved slightly, depending on the balanced dataset. More notably, when examining the performance on the non-balanced datasets, the performance showed significant improvement with TE, indicating that the model made better use of external data. As a result, the best global model performance was achieved when using CoNeTTE trained with CL balanced data.

\begin{table*}[htb]
    \centering
    \caption{AAC results on AC and CL testing subsets. A balanced dataset represents 50\% of the examples seen during an epoch to focus training on that specific dataset. WC* denotes the WavCaps dataset without overlapping sources with AC and CL training, validation and testing subsets, and without audio lasting for more than 30 seconds.}

    \begin{tabular}{ccc|c|ccc|ccc}
        \hline
        \multirow{2}{*}{\textbf{System}} & \multirow{2}{*}{\textbf{TE}} & \multirow{2}{*}{\textbf{Training data}} & \multirow{2}{*}{\textbf{Balanced data}} & \multicolumn{3}{c}{\textbf{AC-test}} & \multicolumn{3}{c}{\textbf{CL-test}} \\ 
        & & & & \textbf{SPIDEr} & \textbf{FENSE} & \textbf{\#Words} & \textbf{SPIDEr} & \textbf{FENSE} & \textbf{\#Words} \\ \hline
        CNext-trans & No & AC & N/A & .466 & \textbf{.633} & 396.4 & .146 & .465 & 401.4 \\
        CNext-trans & No & AC+CL+MA+WC* & AC & .456 & .627 & \textbf{396.8} & .191 & .484 & \first{472.4} \\
        CoNeTTE & Yes & AC+CL+MA+WC* & AC & \textbf{.468} & .630 & 379.8 & \first{.252} & \first{.498} & 467.4 \\
        \hdashline
        CNext-trans & No & CL & N/A & .231 & .521 & \first{525.2} & .301 & .516 & 628.2 \\
        CNext-trans & No & AC+CL+MA+WC* & CL & .364 & .591 & 376.0 & .295 & .510 & 589.2 \\
        CoNeTTE & Yes & AC+CL+MA+WC* & CL & \first{.441} & \first{.609} & {331.2} & \textbf{.305} & \textbf{.517} & \textbf{639.0} \\
        \hline
    \end{tabular}
    \label{tab_both_results}
\end{table*}


\subsection{Does Task Embedding change form or content, or both?}

\begin{table}[htb]
    \centering
    \caption{Captions generated with two different TEs with their corresponding references on the audio file ID \mytexttt{35b9BSmN5JM}.}
    \label{tab_te_examples_1}

    \begin{tabular}{clcc}
        \hline
        \colname{Task} & \colname{Candidates} & \colname{SPIDEr} & \colname{FENSE} \\
        \hline
        AC & A vehicle engine idling and revving & \first{.283} & \first{.592} \\
        \hdashline
        \multirow{2}{*}{CL} & An engine is idling and revving up & \multirow{2}{*}{.249} & \multirow{2}{*}{.577} \\
        & and down & & \\
        \hline
        \multicolumn{4}{l}{\colname{References}} \\
        \hline
        \multicolumn{4}{l}{Loud vibrating followed by revving} \\
        \multicolumn{4}{l}{Truck in idle mode, door closing, engine revving and accelerating} \\
        \multicolumn{4}{l}{A wooden thud as an idle car engine runs then accelerates} \\
        \multicolumn{4}{l}{A motor vehicle accelerates and revs} \\
        \multicolumn{4}{l}{An engine running} \\
        \hline
    \end{tabular}
\end{table}

\begin{table}[htb]
    \centering
    \caption{Captions generated with two different TEs with their corresponding references on the audio file \mytexttt{Diving Bell 1.wav}.}
    \label{tab_te_examples_2}

    \begin{tabular}{clcc}
        \hline
        \colname{Task} & \colname{Candidates} & \colname{SPIDEr} & \colname{FENSE} \\
        \hline
        AC & A musical instrument is playing a note & {.037} & {.325} \\
        \hdashline
        \multirow{2}{*}{CL} & A gong is struck and echoes in a steady & \multirow{2}{*}{\first{.198}} & \multirow{2}{*}{\first{.588}} \\
        & rhythm & & \\
        \hline
        \multicolumn{4}{l}{\colname{References}} \\
        \hline
        \multicolumn{4}{l}{A bell is struck by a mallet, and the noise resonates for some time.} \\
        \multicolumn{4}{l}{A heavy chime is struck and rings loudly at an even tone.} \\
        \multicolumn{4}{l}{A mallet strikes a bell and the sound resonates for a time.} \\
        \multicolumn{4}{l}{A single bell is sounded and its reverberations felt all around.} \\
        \multicolumn{4}{l}{single bell sound followed by its vibration sound} \\
        \hline
    \end{tabular}
\end{table}

To assess the ability of CoNeTTE to produce captions in different writing styles, we tested its ability to generate captions on the CL testing subset with two TE tokens: the CL and the AC ones. Specifically, Tables~\ref{tab_te_examples_1} and \ref{tab_te_examples_2} present various outputs from our model with these two different TEs. These examples demonstrate that the model generates different captions depending on the TE provided. Additionally, we report the SPIDEr, FENSE, vocabulary size, and average sentence length in Table~\ref{tab_te_vocab_sent_len} for both TEs and testing subsets. For instance, on the CL test subset, the sentence length decreased from 10.8 to 7.2, and the vocabulary size dropped from 639.0 to 412.2 words when using the CL TE versus the AC TE. As shown by the decrease in the SPIDEr score from .305 to .227 when using the AC TE, there was a slight drop in performance when using the wrong TE for a dataset. However, despite these changes, we calculated the SPIDEr between the two different outputs and found it to be relatively high at 1.148, indicating that the sentences still contain similar content. 

\begin{table}[htb]
    \caption{Vocabulary and sentence average sizes on AC and CL test subsets, with AC and CL TEs and the CoNeTTE model trained with CL balanced data.}
    \label{tab_te_vocab_sent_len}
    \centering

    \begin{tabular}{cccccc}
        \hline
        \colname{Test} & \colname{Task} & \colname{SPIDEr} & \colname{FENSE} & \colname{\#Words} & \colname{\#Sent} \\
        \hline
        AC & AC & .441 & .609 & 331.2 & 7.5 \\
        AC & CL & .307 & .576 & 517.2 & 10.8 \\
        \hdashline
        CL & AC & .227 & .493 & 412.2 & 7.2 \\
        CL & CL & .305 & .517 & 639.0 & 10.8 \\
        \hline
    \end{tabular}
\end{table}

To further explore the difference between the different TE, we reported as supplementary material the distributions of unigrams and trigrams of stemmed words, and of pos-tags in the candidates and references of AC and CL. We observed that the number of prepositions and conjunctions was twice as high when using the CL TE compared to the AC TE on AC and CL test subsets, suggesting that the captions adopt different formulations. The TE also appears to imitate some specific trigrams like \example{followed by a} from AC and \example{in the background} from CL. Moreover, the trigrams corresponding to audio events are not always in the same ranking order when using different TEs, indicating that TEs also have an impact on the nature of the sound events described in the captions.

\section{Conclusions}

In this article, we introduced our novel CNext-trans model and its refined version, CoNeTTE, which demonstrated state-of-the-art results on AudioCaps with 49.5\% SPIDEr and closely approached the state-of-the-art on Clotho with 30.1\% SPIDEr, all while utilizing fewer parameters than existing models. Additionally, we shed light on potential biases when using AudioCaps and showcased that employing an unbiased pretrained audio encoder has an adverse impact on performance, with 46.6\% SPIDEr. 

Notably, in CoNeTTE, we introduced Task Embedding (TE) tags as input to the model, allowing for the combination of several AAC datasets during training. The utilization of TEs effectively addresses the performance gaps observed across AAC datasets by efficiently merging data rather than relying solely on simple concatenation. However, we also uncovered that dataset balancing during training significantly influences the final performance, implying that the one-size-fits-all model still retains some dependency on a specific dataset.

To provide a comprehensive analysis, we presented examples and insights into the impact of TEs on the generated captions, supplemented by word stem and POS tag distributions in the supplementary material. Nevertheless, further investigation in this area would be valuable and intriguing. The combination of AAC datasets warrants additional exploration, either through discovering an optimal balancing policy across datasets or by extending the application of TEs beyond what was used in this study.

\section*{Acknowledgments}
\noindent This work was partially supported by the Agence Nationale de la Recherche the LUDAU (Lightly-supervised and Unsupervised Discovery of Audio Units using Deep Learning) project (ANR-18-CE23-0005-01) and the ANR-3IA Artificial and Natural Intelligence Toulouse Institute. This work was granted access to the HPC resources of IDRIS under the allocation 2022-AD011013739 made by GENCI.




\bibliographystyle{IEEEtran}
\bibliography{refs.bib}


\newcommand\width{0.46\linewidth}

\captionsetup[subfloat]{font=normalsize,labelfont=scriptsize,textfont=scriptsize}

\begin{figure*}[ht]
    \centering
    \subfloat[Distribution of the unigrams in the candidate captions on AC-test with AC TE.]{
        \includegraphics[width=\width]{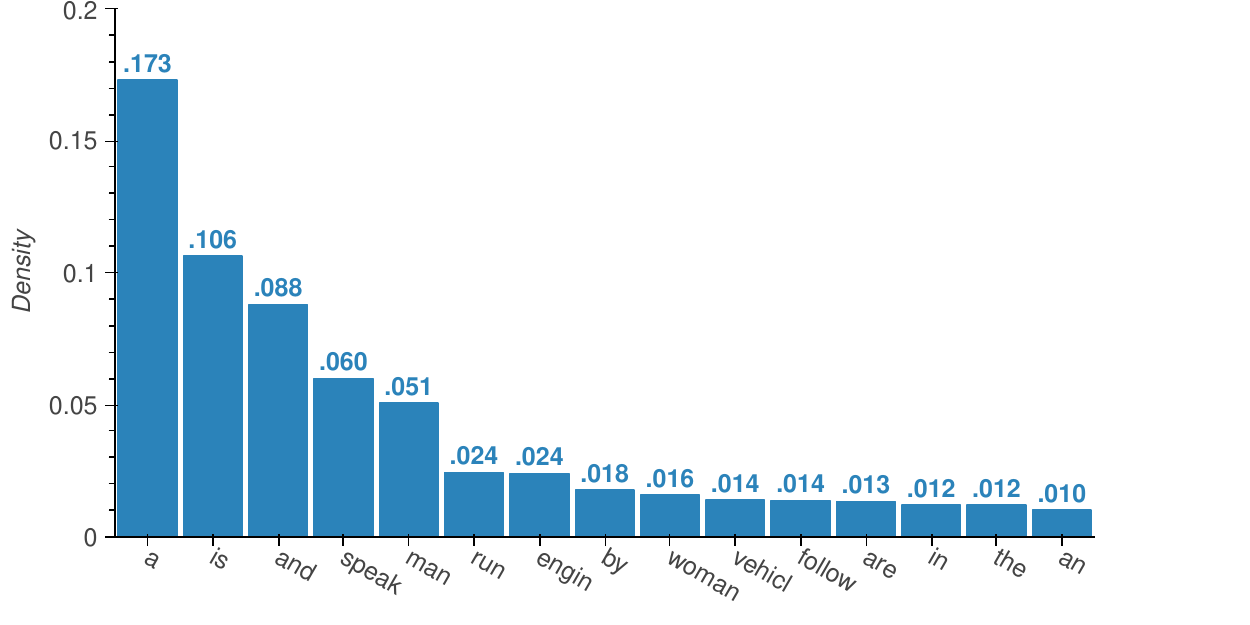}
        \label{fig_1_grams_dset_ac_test_task_ac}
    }\qquad
    \subfloat[Distribution of the unigrams in the candidate captions on AC-test with CL TE.]{
        \includegraphics[width=\width]{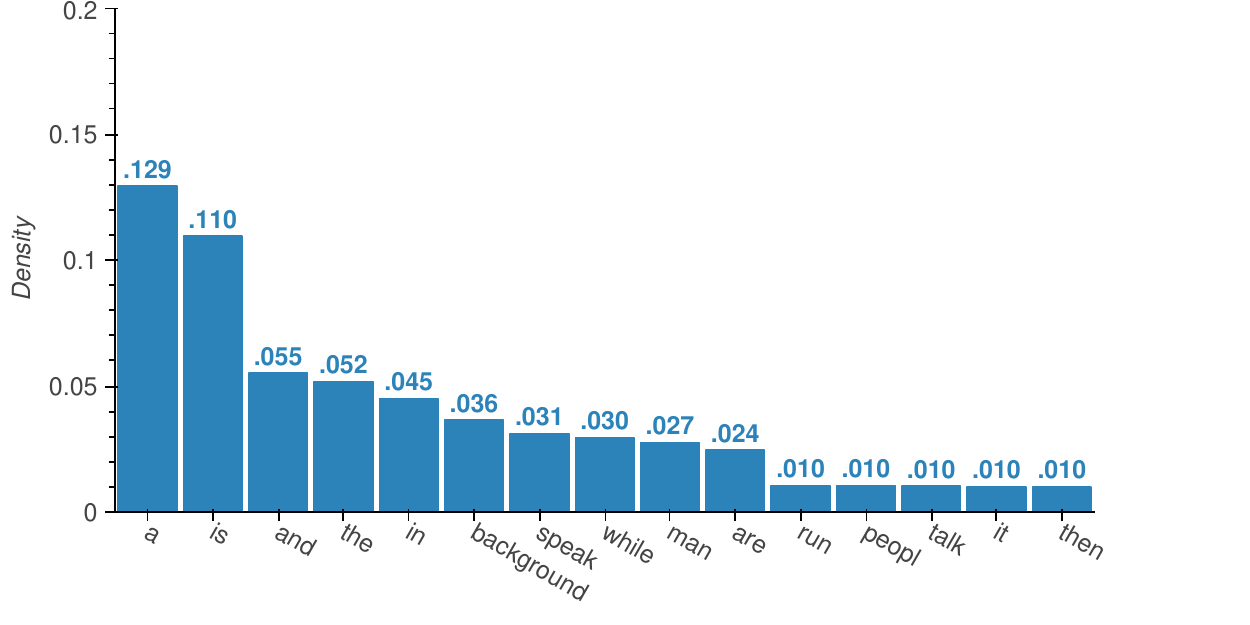}
        \label{fig_1_grams_dset_ac_test_task_cl}
    }\\
    \subfloat[Distribution of the unigrams in the reference captions on AC-train.]{
        \includegraphics[width=\width]{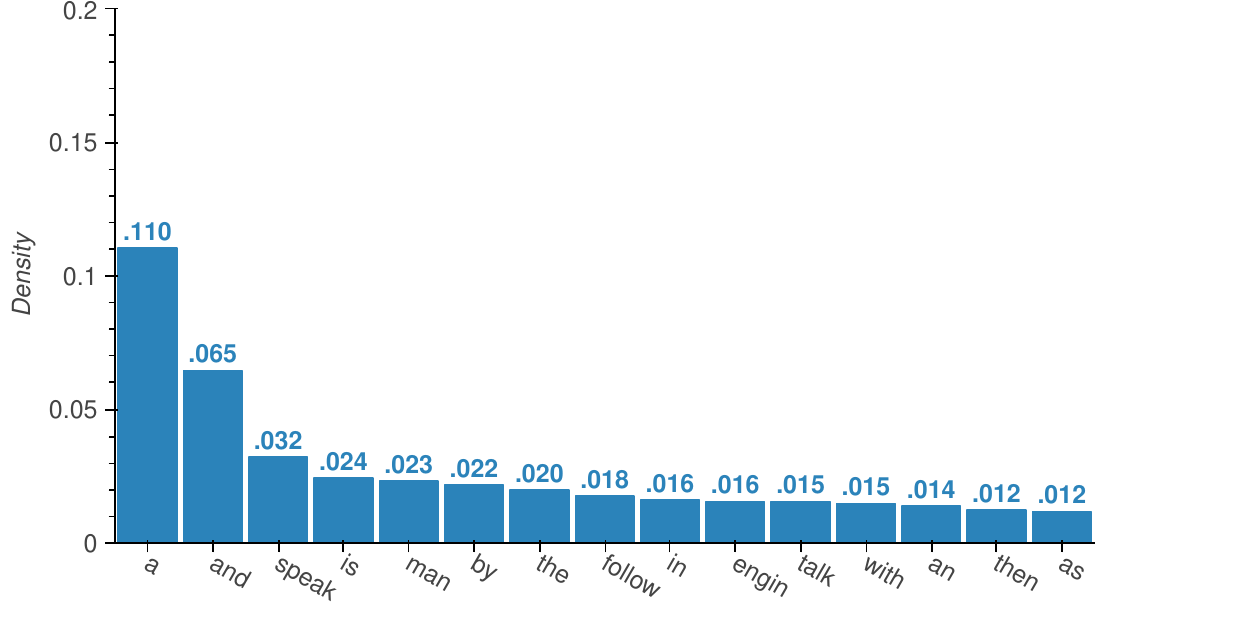}
        \label{fig_1_grams_dset_ac_train}
    }\qquad
    \subfloat[Distribution of references unigrams on AC-test.]{
        \includegraphics[width=\width]{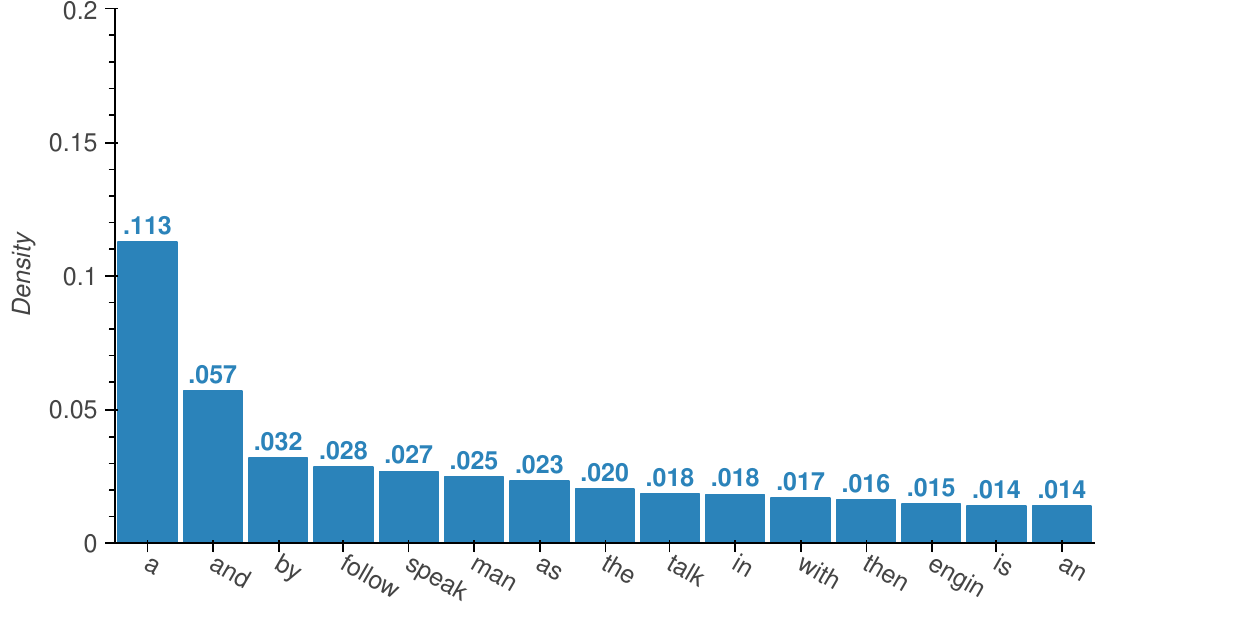}
        \label{fig_1_grams_dset_ac_test}
    }
    \caption{Distributions of unigrams on AC-test and AC-train.}
\end{figure*}

\begin{figure*}[hb]
    \centering
    \subfloat[Distribution of candidates unigrams on CL-eval with AC TE.]{
        \includegraphics[width=\width]{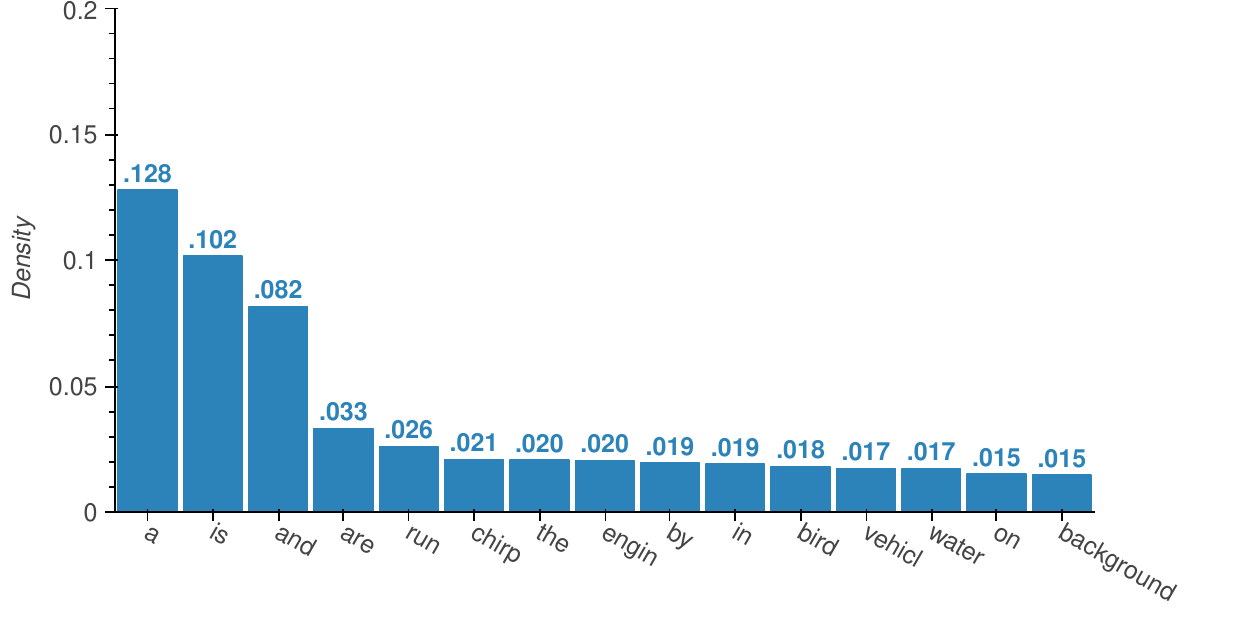}
        \label{fig_1_grams_dset_cl_eval_task_ac}
    }\qquad
    \subfloat[Distribution of candidates unigrams on CL-eval with CL TE.]{
        \includegraphics[width=\width]{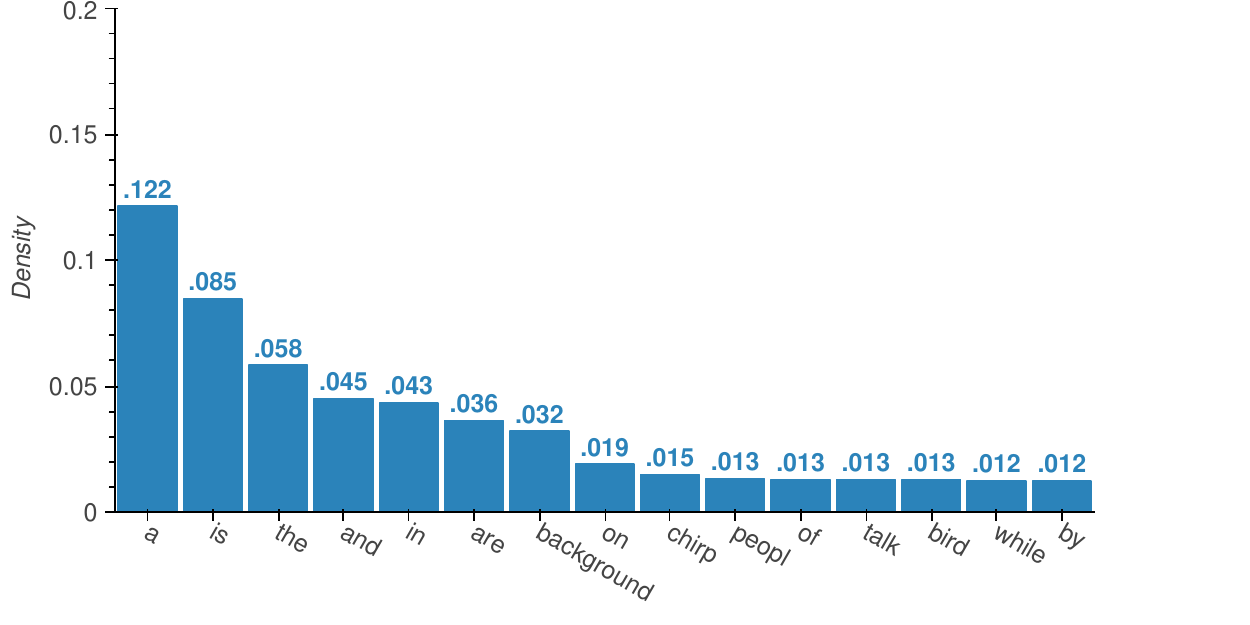}
        \label{fig_1_grams_dset_cl_eval_task_cl}
    }\\
    \subfloat[Distribution of references unigrams on CL-dev.]{
        \includegraphics[width=\width]{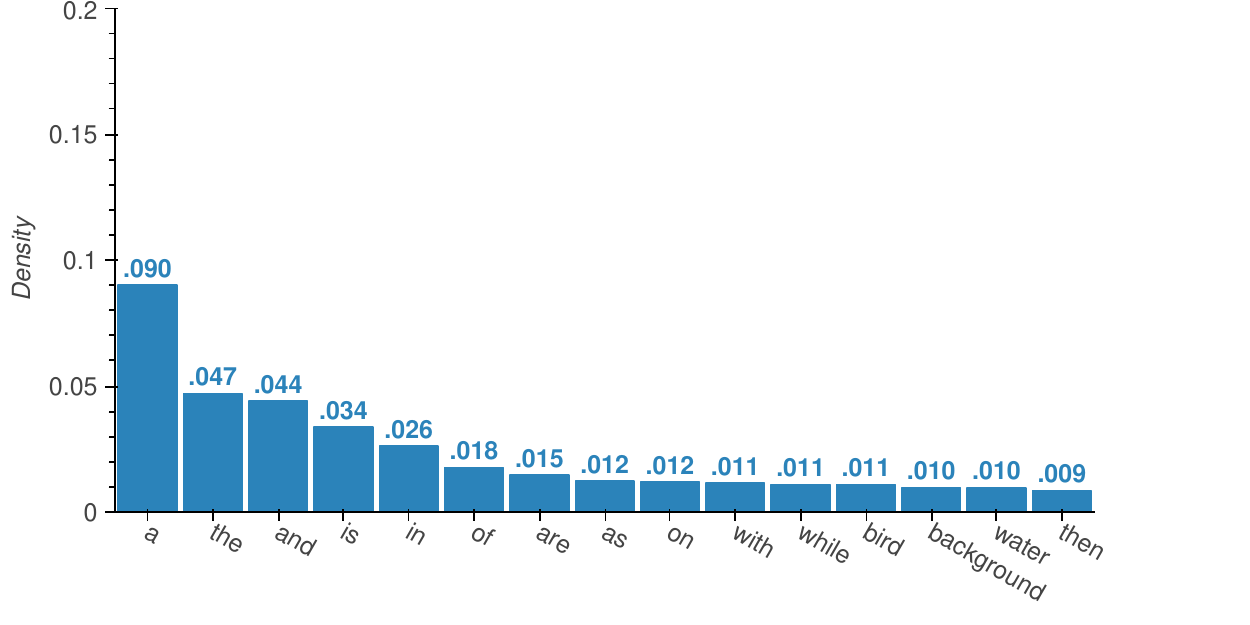}
        \label{fig_1_grams_dset_cl_dev}
    }\qquad
    \subfloat[Distribution of references unigrams on CL-eval.]{
        \includegraphics[width=\width]{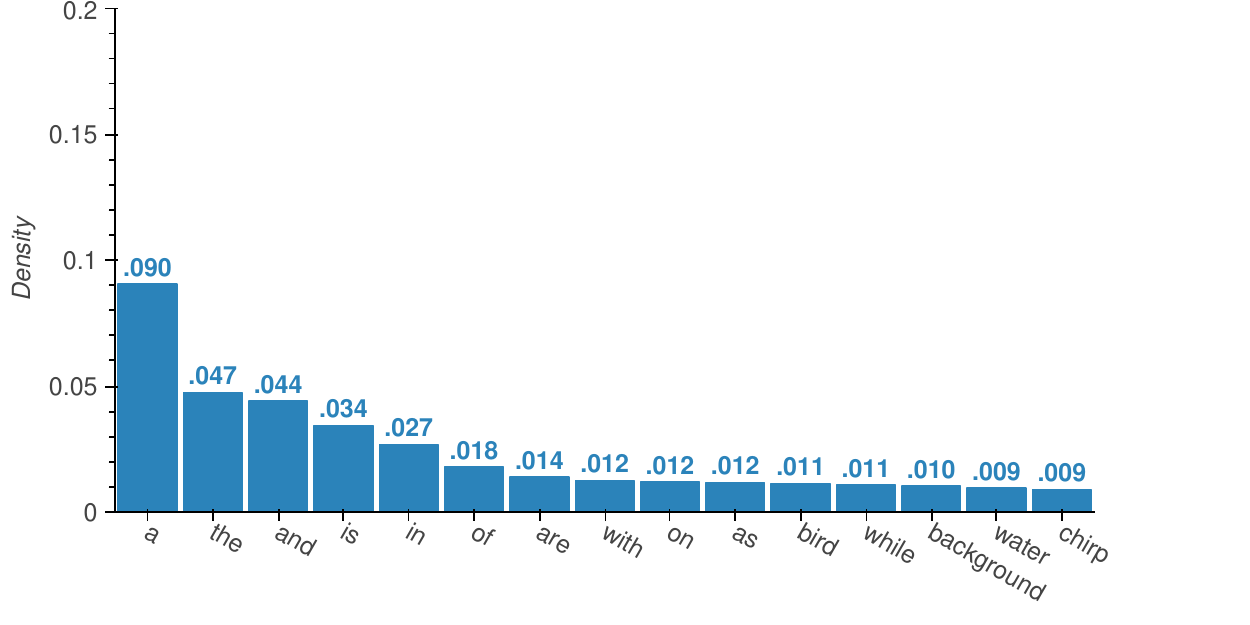}
        \label{fig_1_grams_dset_cl_eval}
    }
    \caption{Distributions of unigrams on CL-eval and CL-dev.}
\end{figure*}

\begin{figure*}[hb]
    \centering
    \subfloat[Distribution of candidates trigrams on AC-test with AC TE.]{
        \includegraphics[width=\width]{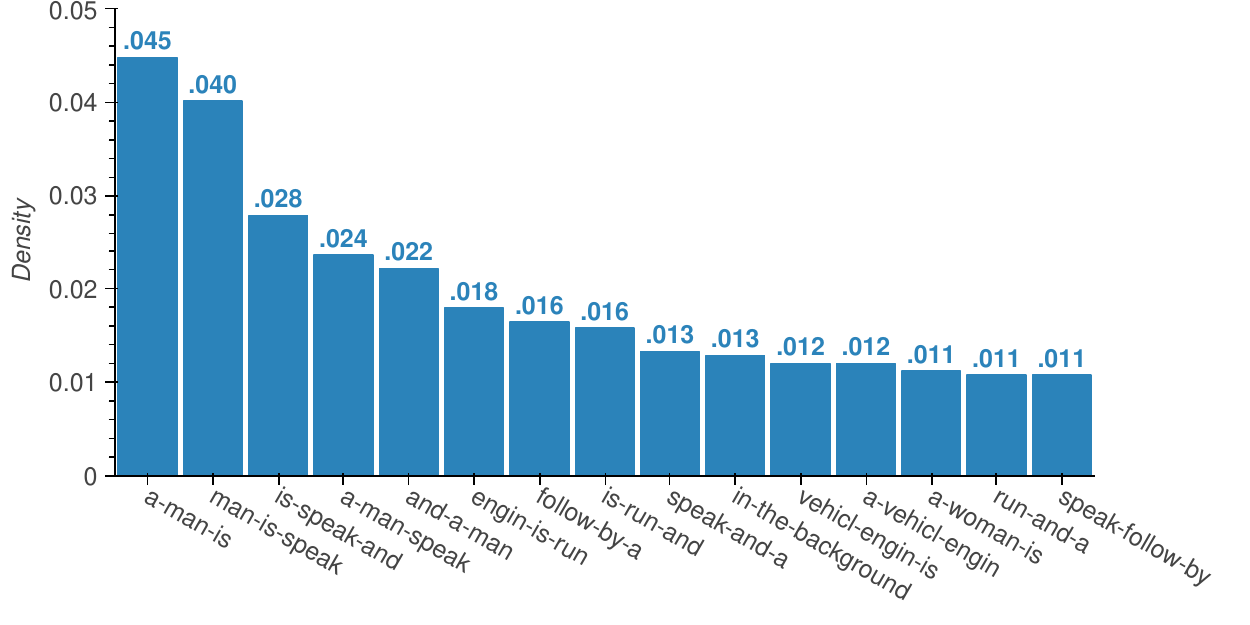}
        \label{fig_3_grams_dset_ac_test_task_ac}
    }\qquad
    \subfloat[Distribution of candidates trigrams on AC-test with CL TE.]{
        \includegraphics[width=\width]{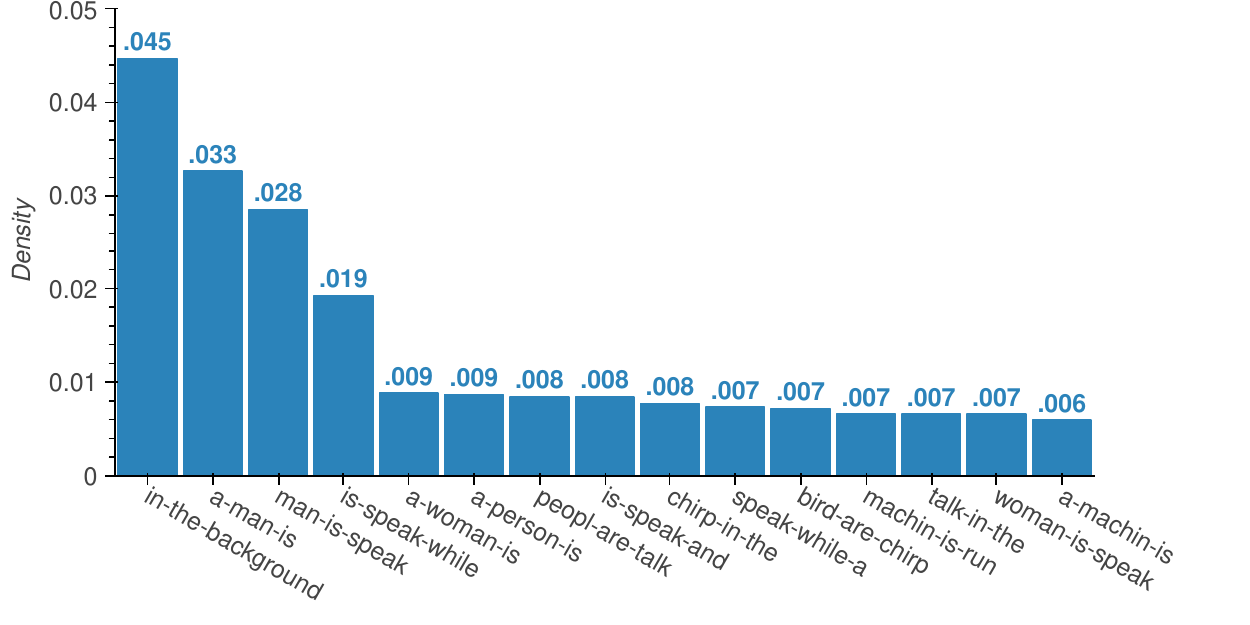}
        \label{fig_3_grams_dset_ac_test_task_cl}
    }\\
    \subfloat[Distribution of references trigrams on AC-train.]{
        \includegraphics[width=\width]{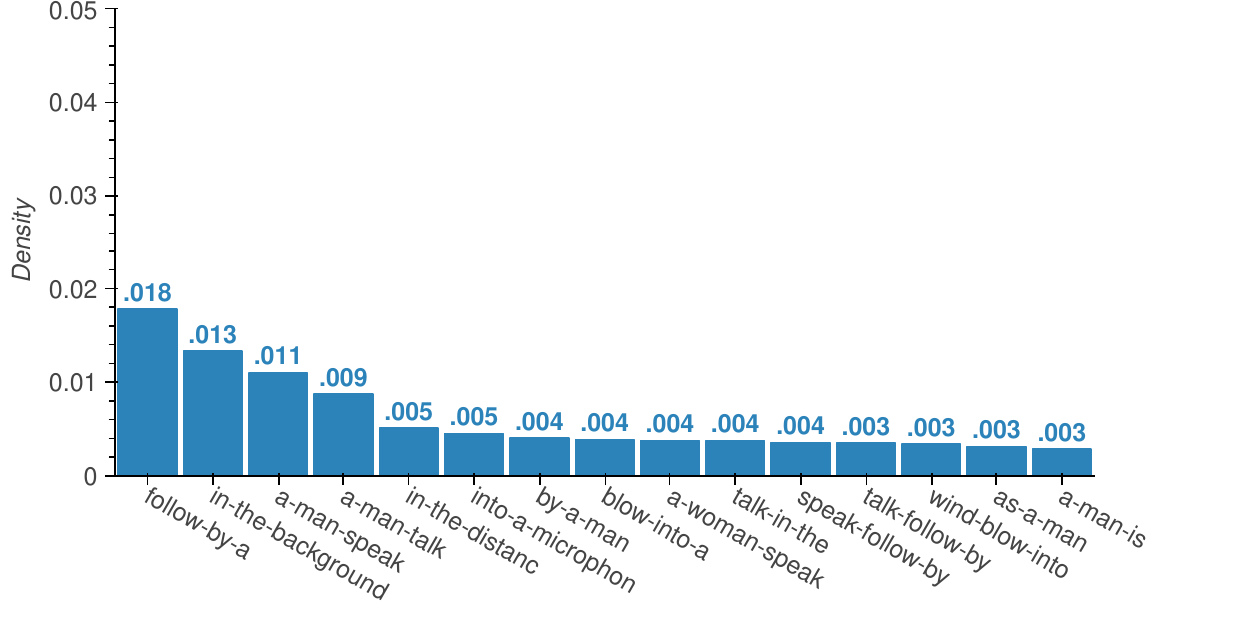}
        \label{fig_3_grams_dset_ac_test}
    }\qquad
    \subfloat[Distribution of references trigrams on AC-test.]{
        \includegraphics[width=\width]{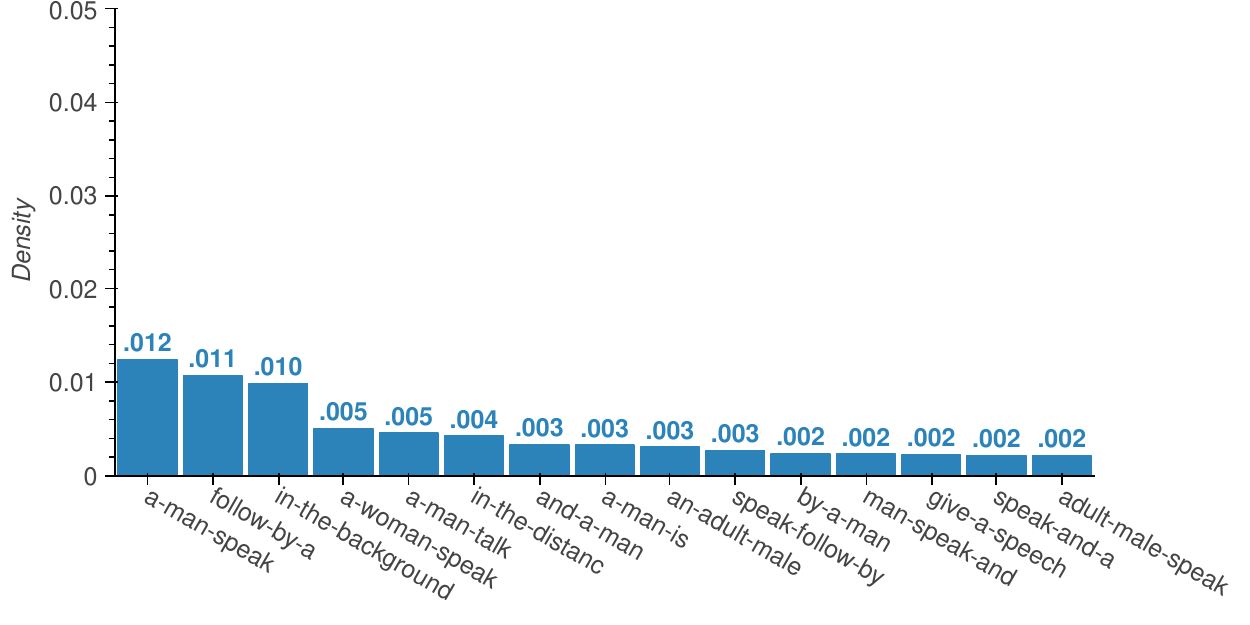}
        \label{fig_3_grams_dset_ac_train}
    }
    \caption{Distributions of trigrams on AC-test and AC-train.}
\end{figure*}

\begin{figure*}[hb]
    \centering
    \subfloat[Distribution of candidates trigrams on CL-eval with AC TE.]{
        \includegraphics[width=\width]{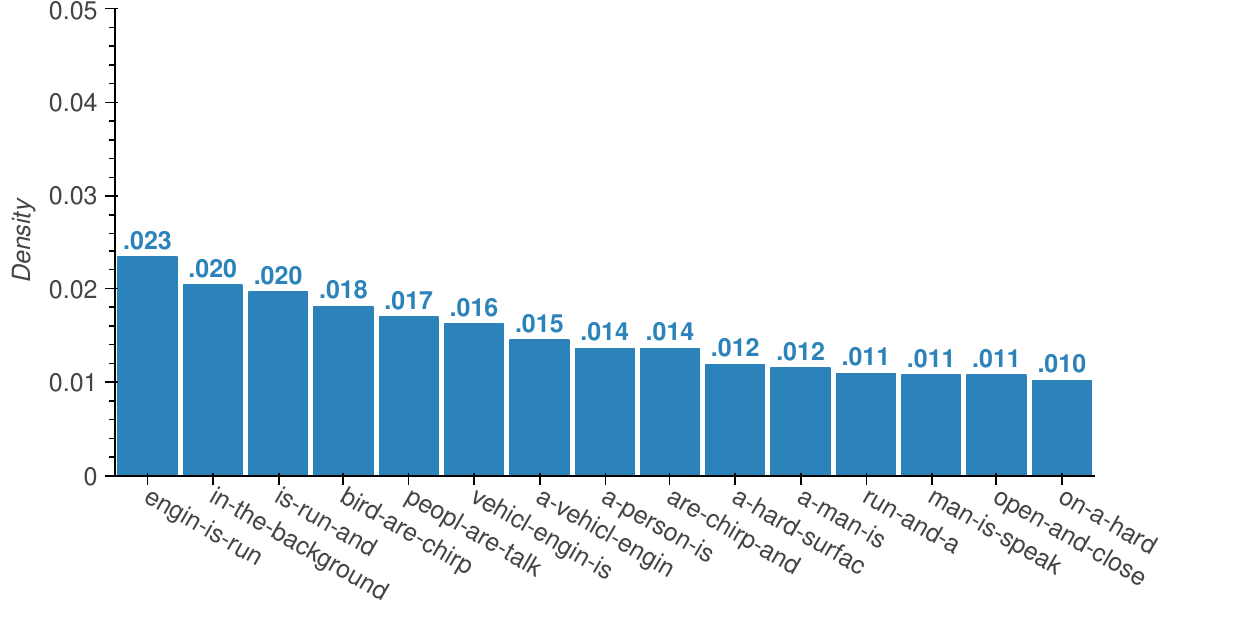}
        \label{fig_3_grams_dset_cl_eval_task_ac}
    }\qquad
    \subfloat[Distribution of candidates trigrams on CL-eval with CL TE.]{
        \includegraphics[width=\width]{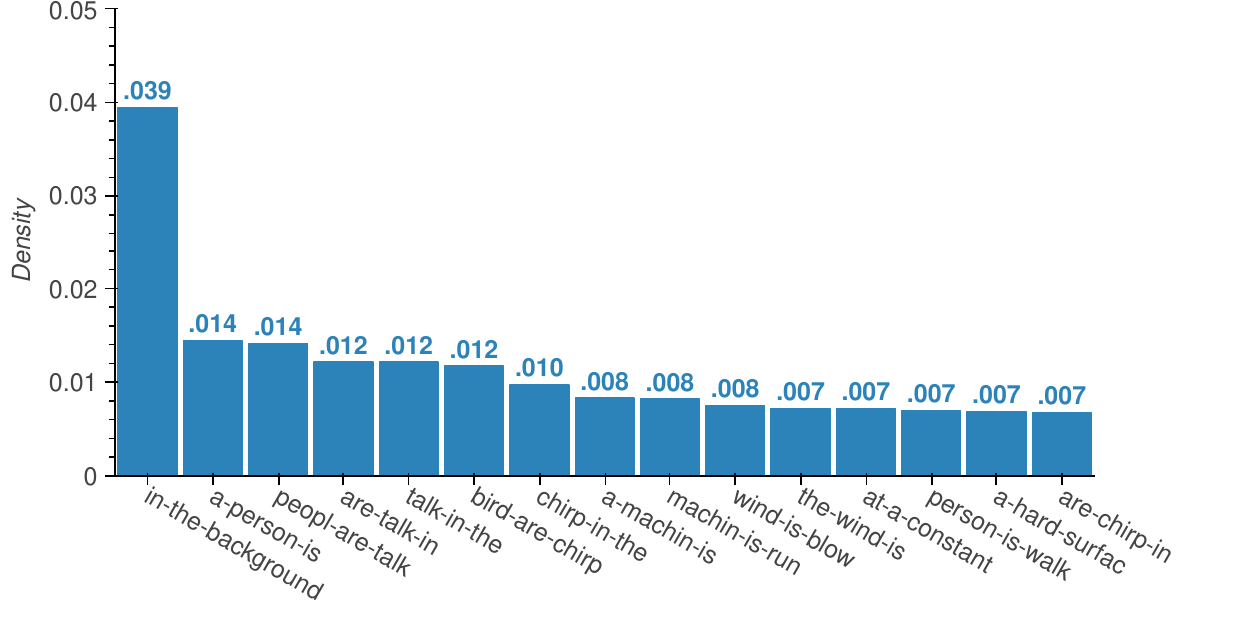}
        \label{fig_3_grams_dset_cl_eval_task_cl}
    }\\
    \subfloat[Distribution of references trigrams on CL-dev.]{
        \includegraphics[width=\width]{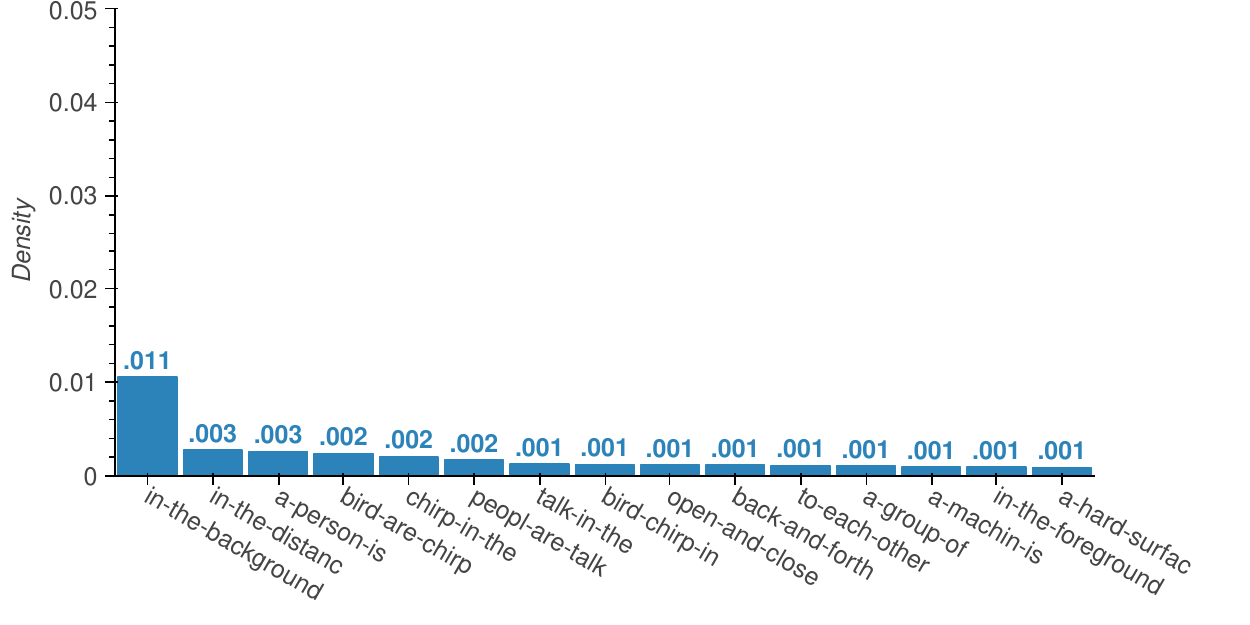}
        \label{fig_3_grams_dset_cl_dev}
    }\qquad
    \subfloat[Distribution of references trigrams on CL-eval.]{
        \includegraphics[width=\width]{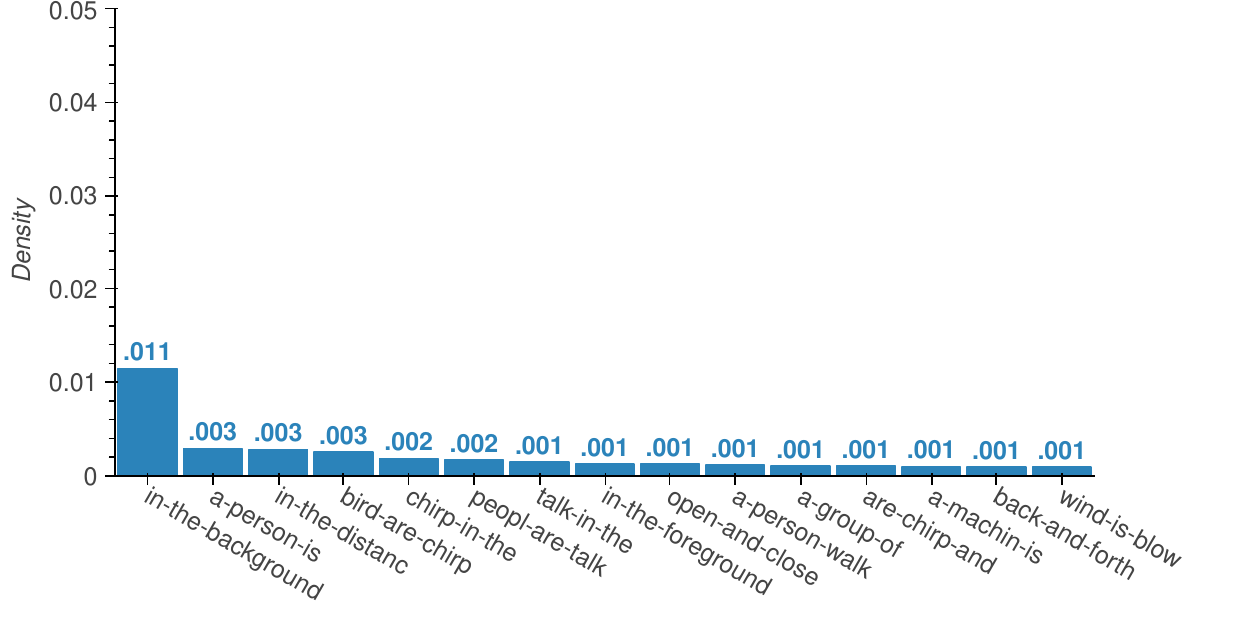}
        \label{fig_3_grams_dset_cl_eval}
    }
    \caption{Distributions of trigrams on CL-eval and CL-dev.}
\end{figure*}

\begin{figure*}[hb]
    \centering
    \subfloat[Distribution of candidates POS-TAGs on AC-test with AC TE.]{
        \includegraphics[width=\width]{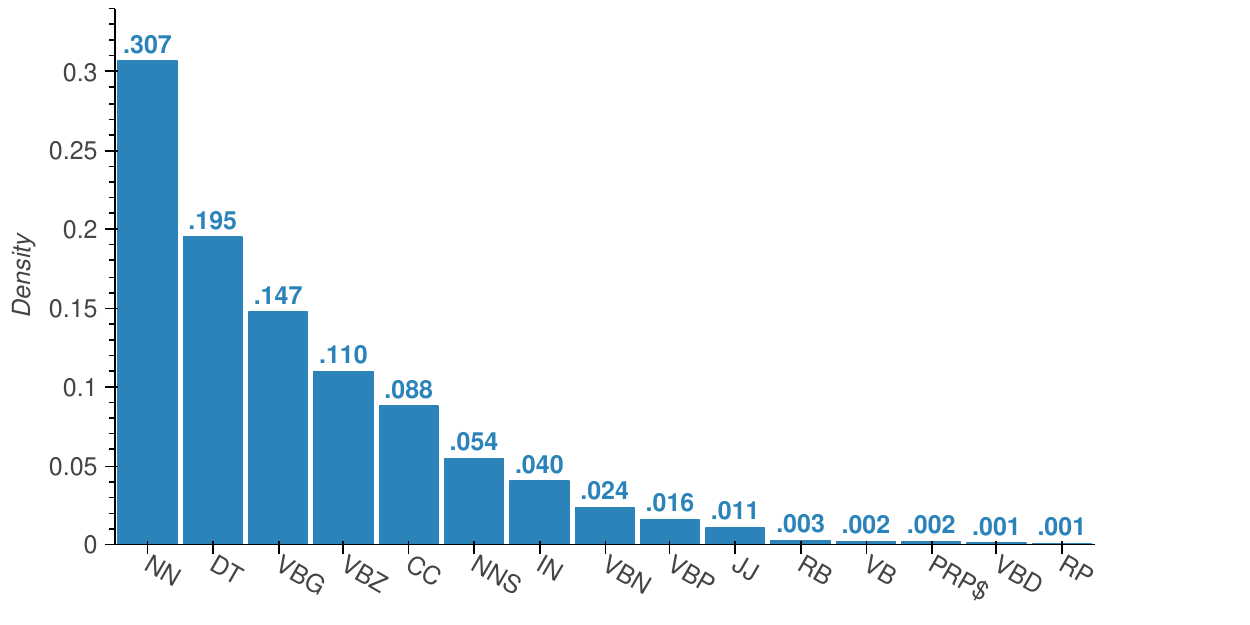}
        \label{fig_ptags_dset_ac_test_task_ac}
    }\qquad
    \subfloat[Distribution of candidates POS-TAGs on AC-test with CL TE.]{
        \includegraphics[width=\width]{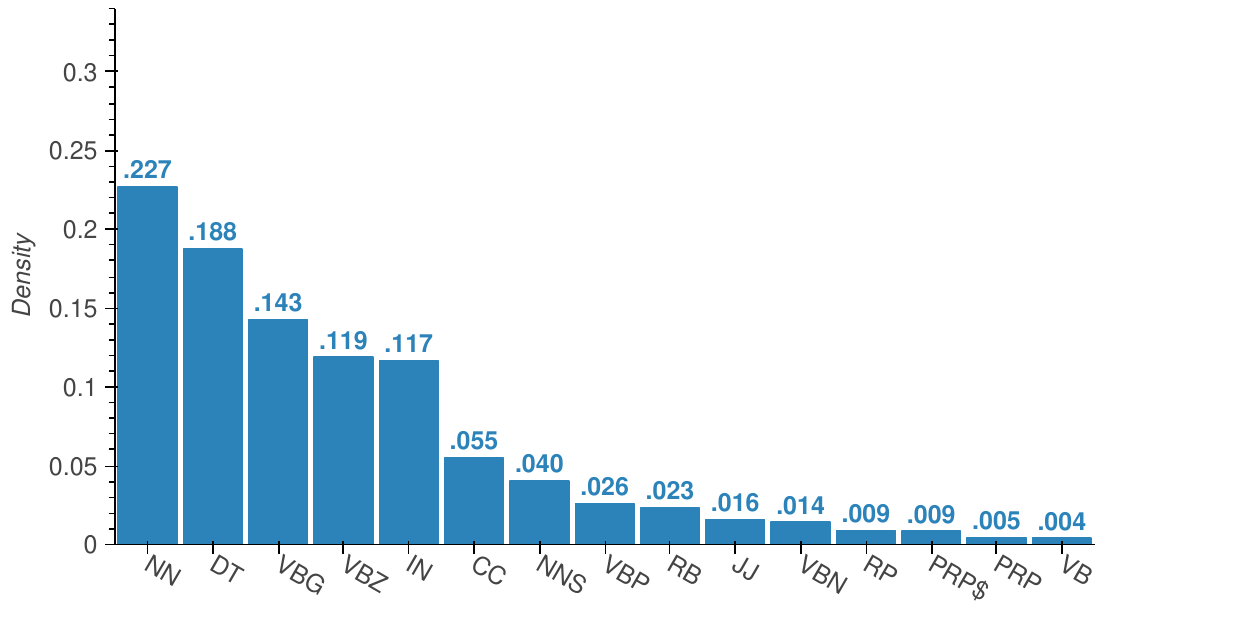}
        \label{fig_ptags_dset_ac_test_task_cl}
    }\\
    \subfloat[Distribution of references POS-TAGs on AC-train.]{
        \includegraphics[width=\width]{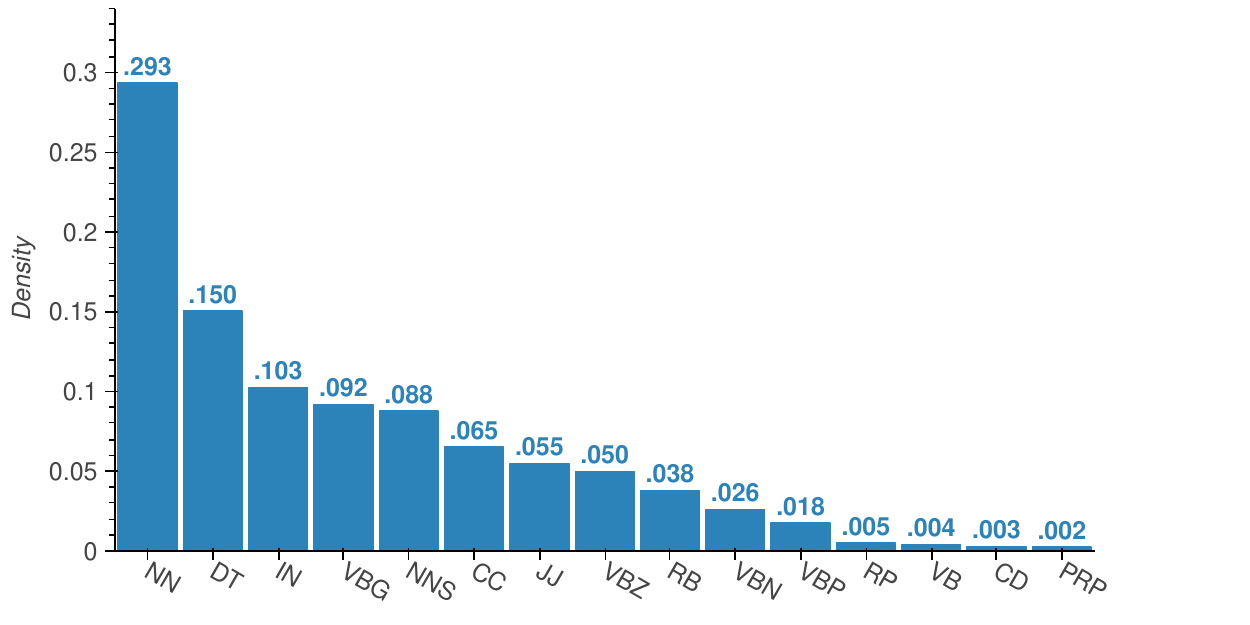}
        \label{fig_ptags_dset_ac_train}
    }\qquad
    \subfloat[Distribution of references POS-TAGs on AC-test.]{
        \includegraphics[width=\width]{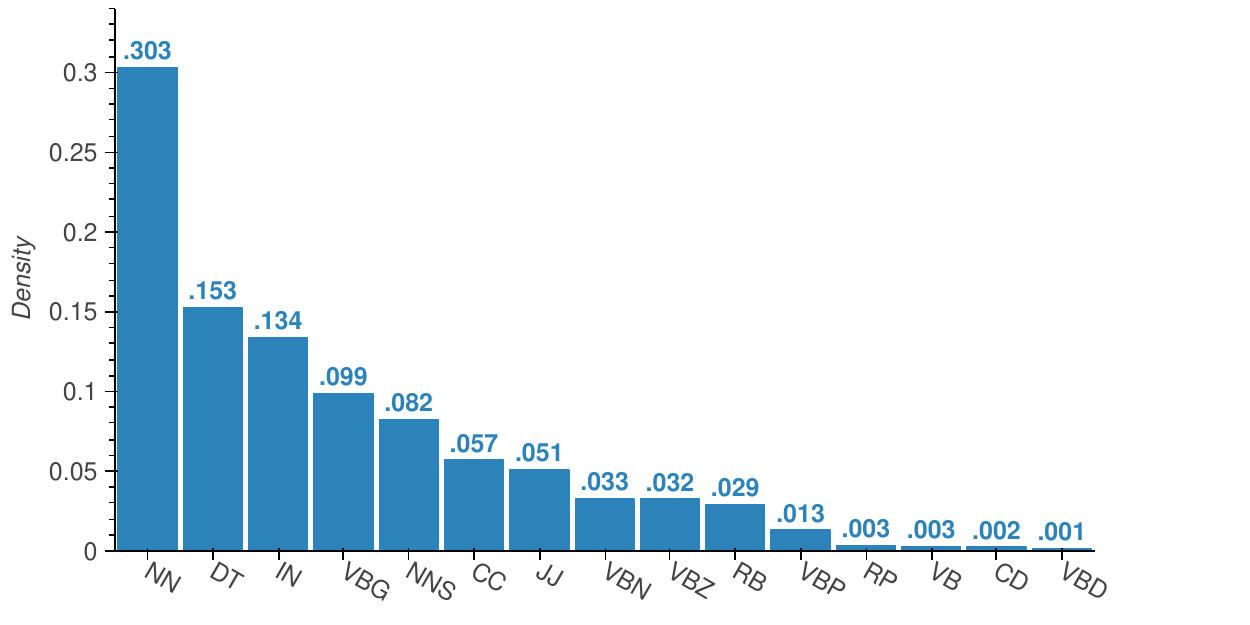}
        \label{fig_ptags_dset_ac_test}
    }
    \caption{Distributions of POS-TAGs on AC-test and AC-train.}
\end{figure*}

\begin{figure*}[hb]
    \centering
    \subfloat[Distribution of candidates POS-TAGs on CL-eval with AC TE.]{
        \includegraphics[width=\width]{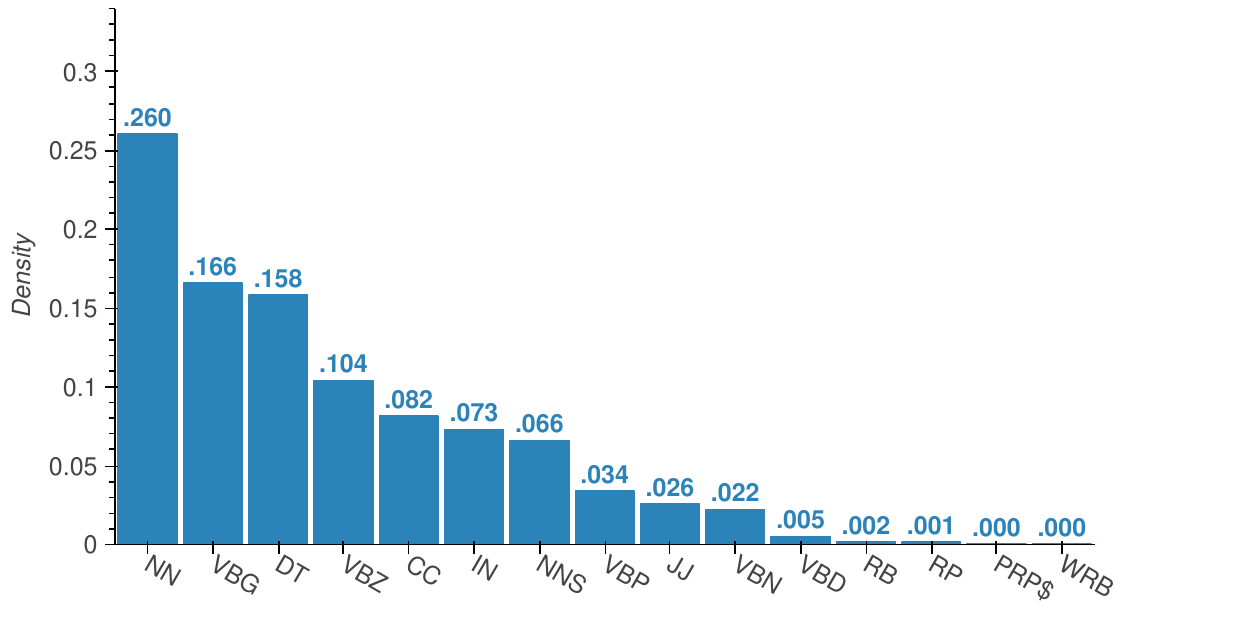}
        \label{fig_ptags_dset_cl_eval_task_ac}
    }\qquad
    \subfloat[Distribution of candidates POS-TAGs on CL-eval with CL TE.]{
        \includegraphics[width=\width]{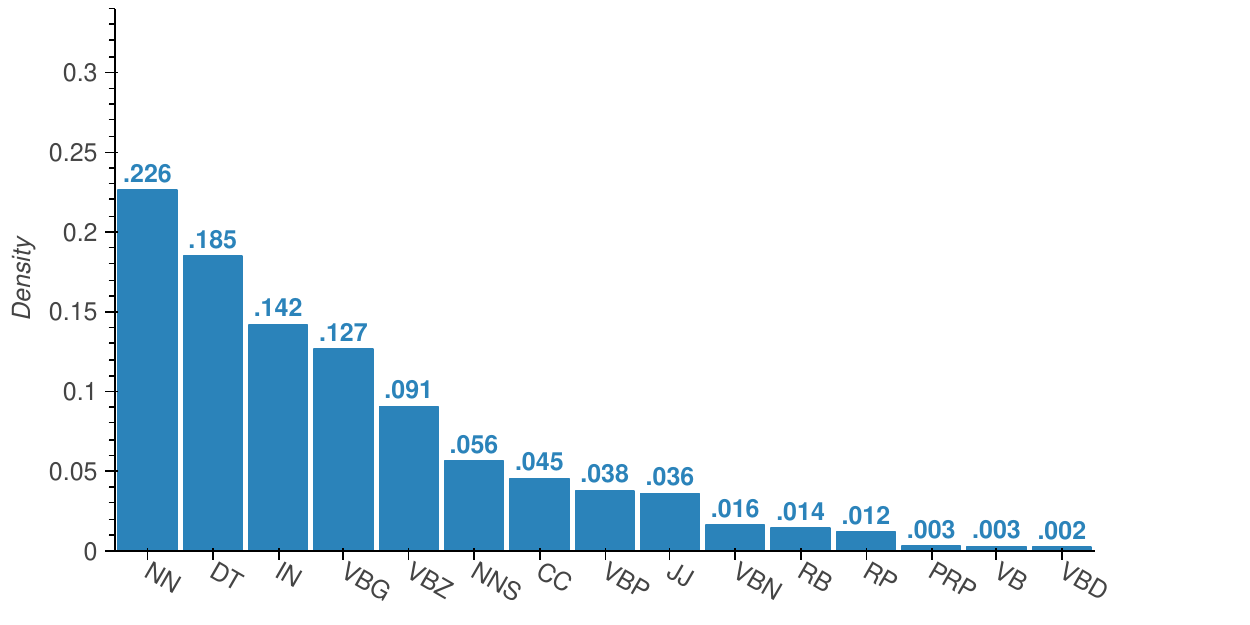}
        \label{fig_ptags_dset_cl_eval_task_cl}
    }\\
    \subfloat[Distribution of references POS-TAGs on CL-dev.]{
        \includegraphics[width=\width]{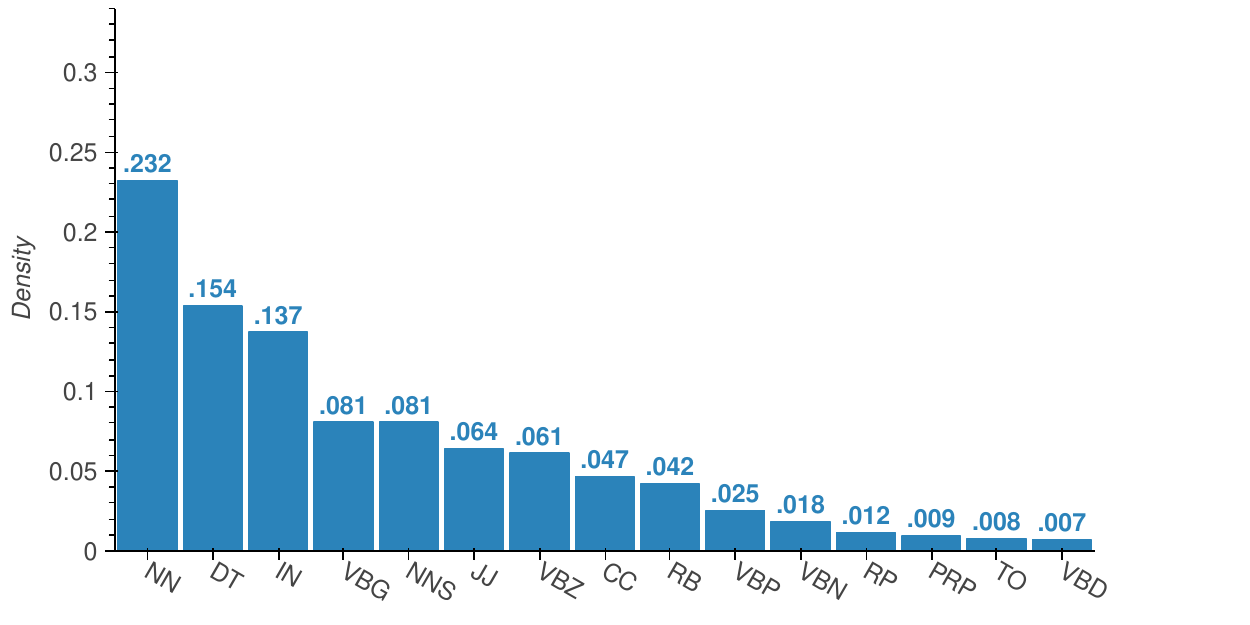}
        \label{fig_ptags_dset_cl_dev}
    }\qquad
    \subfloat[Distribution of references POS-TAGs on CL-eval.]{
        \includegraphics[width=\width]{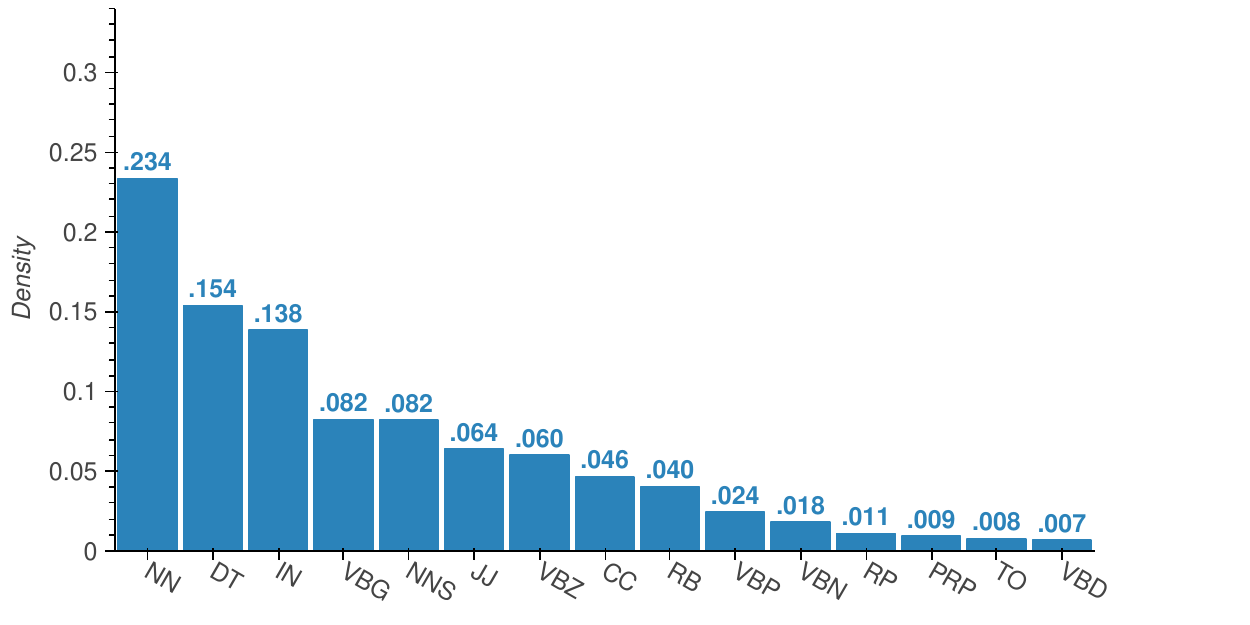}
        \label{fig_ptags_dset_cl_eval}
    }
    \caption{Distributions of POS-TAGs on CL-eval and CL-dev.}
\end{figure*}


\end{document}